\documentclass[natbib]{svjour3}                     
\smartqed  
\usepackage{times}
\usepackage{graphicx}
\usepackage{epsf}
 \usepackage{aps-bibstyle}  
\renewcommand{\etal}{{\it et al.}}
\begin{document}

\title{Magnetic Fields in Astrophysical Jets: From Launch to Termination
}


\author{Ralph~E.~Pudritz      \and
        Martin~J.~Hardcastle \and Denise~C.~Gabuzda 
}


\institute{Ralph Pudritz \at
              Dept. of Physics and Astronomy, McMaster University, Hamilton, ON L8S 4M1, Canada\\
              \email{pudritz@mcmaster.ca}           
           \and
           Martin Hardcastle \at
              School of Physics, Astronomy \& Mathematics, University
              of Hertfordshire, College Lane, Hatfield AL10 9AB, UK\\
              \email{m.j.hardcastle@herts.ac.uk} 
              \and
              Denise Gabuzda \at
              Physics Department, University College Cork, Republic of
              Ireland\\
              \email{d.gabuzda@ucc.ie}
}

\date{Received: date / Accepted: date}

\maketitle

\begin{abstract}

Long-lived, stable jets are observed in a wide variety of systems,
from protostars, through Galactic compact objects to active galactic
nuclei (AGN). Magnetic fields play a central role in launching,
accelerating, and collimating the jets through various media. The
termination of jets in molecular clouds or the interstellar medium
deposits enormous amounts of mechanical energy and momentum, and their
interactions with the external medium, as well, in many cases, as the
radiation processes by which they are observed, are intimately
connected with the magnetic fields they carry. This review focuses on
the properties and structures of magnetic fields in long-lived jets,
from their launch from rotating magnetized young stars, black holes,
and their accretion discs, to termination and beyond. We compare the
results of theory, numerical simulations, and observations of these
diverse systems and address similarities and differences between
relativistic and non-relativistic jets in protostellar versus AGN
systems. On the observational side, we focus primarily on jets driven
by AGN because of the strong observational constraints on their
magnetic field properties, and we discuss the links between the
physics of these jets on all scales.

\keywords{magnetic fields \and jets \and polarization \and radio
  continuum: galaxies \and X-rays: galaxies}
\end{abstract}

\section{Introduction}
\label{intro}

Astrophysical jets have been discovered in a very wide range of
physical systems, from regions of star formation, to gamma-ray
bursters and stellar black holes, on up in scale and energy to the
nuclei of giant radio galaxies and quasars. They are often associated with
accretion discs and their launch, internal dynamics, and impact upon
their surroundings are of ongoing importance and significance over a
whole range of astrophysical phenomena --- from the parsec (pc) scales
that characterize the impact of jets from young stellar objects within
their natal molecular clouds, to the megaparsec (Mpc) scales of giant
clusters of galaxies whose intercluster media are significantly
stirred and heated by jets emanating from the cores of active galaxies
\citep[e.g.,][]{Dunn/Fabian:2006}. On stellar scales, jets from young
stellar objects (YSOs) span the breadth of the stellar mass spectrum:
from massive stars such as W33A which shows a hot fast jet when
observed in Br $\gamma$ lines as well as an underlying accretion disc
\citep{Davies/etal:2010}, to the well known solar-mass systems
such as classical T-Tauri stars (TTS) \citep[see][for a
  review]{Ray/etal:2007}, and on down into the brown dwarfs
\citep[e.g.,][]{Whelan/etal:2009}. There is good evidence that all of
the protostellar systems are associated with discs. Stellar mass black
holes, otherwise known as microquasars \citep{Mirabel:2005} are
also very active sources of relativistic jets.   
On extragalactic scales, jets
are observed to be associated with accretion on to supermassive black
holes (up to a few billion solar masses) in the centres of active
galaxies \citep{Blandford:2001}. Models for gamma-ray
bursts (GRBs), which are also located at cosmological distances,
focus on either the collapse of massive stars or the merger
of two compact stellar objects to form a black hole in a process that 
releases about a solar mass worth of gravitational potential
energy and which generates a highly relativistic magnetized jet \citep{Meszaros:2012}.

Given the vast range in energies, scales, and details spanned by these
diverse systems, it is intriguing that accreting systems seen in
nature make such extensive and ubiquitous use of energetic outflows as
part of their evolution. Is there a universal mechanism at work that
almost guarantees that jets will be present? What is the basic
physical mechanism that creates and drives jets across all of these
scales? What role do they play in the evolution of such accreting
compact systems? The fact that all these systems share common physical
components --- central gravitating sources, accretion discs, and
magnetic fields --- provides the unifying foundation on which
theoretical models are constructed and observationally tested.

Magnetic fields play a critical role in creating, driving, and
directing the evolution of jets. Indeed, the successful existing
models of jets in these diverse systems invoke magnetic fields
threading central objects such as gaseous accretion discs or rapidly
rotating central objects themselves \citep[see reviews by,
  e.g.,][]{Shu/etal:2000,Pudritz/etal:2007}. The release of
gravitational potential energy during accretion is a key power source
for most of these jets. The great utility of magnetic torques is that
they can readily tap into the shear energy available in the accretion
discs that form as a consequence of gravitational collapse. Models of
magnetized outflows from accretion discs in AGN, developed by
\citet{BlandfordPayne1982}, were applied soon after to understanding
jets in protostellar systems \citep{Pudritz/Norman:1983,
  Pudritz/Norman:1986}. These theories envisage that material from a
disc, in being forced to co-rotate with a threading magnetic field, is
flung outwards along sufficiently inclined, open field lines. The gas
continues to accelerate as long as the field is strong enough to
maintain co-rotation with the central disc, or rotor. The inertia of
this matter moving along the field line causes the field to wrap up in
a toroidal-like structure --- creating a helical field. It is this
dominant, outer toroidal field component that exerts a pinching force
upon the jet, collimating it towards the axis. A second major
prediction of this basic theory is that jets should be extracting the
angular momentum of their sources, and jet rotation is a key
prediction of the models. Another abundant source of energy resides in
the rotation available in rapidly rotating black holes or pulsars, and
once again it is known that electromagnetic torques can tap rotational
energy efficiently \citep{Blandford/Znajek:1977}. So for rapidly
accreting or spinning objects, large scale magnetic fields appear to
be nature's preferred tool in extracting energy and angular momentum
and directing them in the formation of ordered, magnetically
collimated, outflows.

Observations of crucial properties of jets, such as their densities,
temperatures, velocity structure, and magnetic fields, vary quite
considerably between protostellar and extragalactic systems. In the
former, forbidden line emission (FLER) produced by shocks in the jet
allows one to measure directly everything with the exception of the
magnetic field. On the other hand, extragalactic jets emit synchrotron
radiation which arises from the relativistic electrons that spiral
around magnetic field lines. This radiation allows one directly to
measure the direction of the magnetic field in the jet, and to
estimate its strength.
However, the absence of distinct emission lines from the outflowing
material itself in AGN jets prevents one from measuring the other
crucial physical variables. It is for this reason that protostellar
systems have become such important laboratories for jet studies. Until
2010, these limitations prevented direct observational comparison of
these classes of jets, and, as a result, meant that it was impossible
to test whether or not jets have a universal magnetic field based
mechanism. This all changed with the discovery of synchrotron
radiation from the protostellar jet, HH 80-81
\citep{Carrasco-Gonzalez/etal:2010}. Relativistic electrons in
otherwise non-relativistic jets arise from acceleration in the jet's
terminal shock. The results show that the magnetic structures of
protostellar and AGN jets are similar --- with a field parallel to the
jet axis at the centre and a surrounding helical field towards the
jet edges.

In this paper, we examine the structure and role of magnetic fields in
the physics of jets --- from their launch on discs or compact rotors,
to their dynamics and evolution, and finally to a consideration of how
such strongly magnetized flows affect their environments on these
various scales. We begin our analysis with non-relativistic systems
--- the protostellar jets that are a ubiquitous and vital component of
any star-forming region (Section 2). We also summarize the theory
and simulations of relativistic jets in the context of AGNs.  
(We do not consider the impulsive, short-lived jets in gamma-ray
bursts, although much of the same physics is expected to apply, as the
observational challenges are very different; the reader is referred to
reviews by \cite{Piran2004} and \cite{Granot2007}.) In Section 3 we then turn to
relativistic systems and discuss the measurement of the structure and
dynamics of AGN jets on highly compact (pc) scales, probing the
conditions near at the source of the jets using VLBI. The magnetic
structure of jets on these scales is probably a reflection of their
internal dynamics and not of the shocks which occur throughout the
body of the jets. Indeed, the mounting evidence for helical magnetic
fields in extragalactic jets, taken with the newly discovered evidence
for helical fields in protostellar jets discussed above, strongly
suggest that the physical mechanism of collimation may be the same.
Finally, we go out to the scales of jet termination (Section 4),
focussing on the evidence for magnetic field properties on these
largest scales and exploring the connection between
magnetic fields in jets in the hearts of AGNs with jets at the largest
scales (1 kpc--1 Mpc). We conclude our study with a discussion of
future prospects in this field (Section 5).

\section{Magnetic launching of jets}

\subsection{Observed properties of protostellar jets}

Observations of the launch regions even of protostellar jets face significant
observational challenges. For the nearest regions of star formation
and jets in the Taurus molecular cloud (distance 140 pc), $1''$
resolution corresponds to a physical scale of 140 AU. Theoretical
models discussed below picture a jet acceleration region which is of
the order of several tens of AU ($0.1''$), and the launch region in
the very inner regions of the disc (several AU) that may be an order
of magnitude smaller ($0.01''$). Jets around young stars have been
studied on scales ranging from tens of AU to 5 pc
\citep{Bally/etal:2007}.

{\it Hubble Space Telescope} ({\it HST}) observations of jets from
YSOs have advanced our experimental understanding of jets enormously
over the last decade, probing physical scales down to tens of AU.
There are four basic types of observations that point towards the
strong coupling of jets with underlying accretion discs --- the
correlation of jets with discs, the onion-like velocity structure of
jets, jet rotation, and the strong link between the rates of mass
transport in the jet to the accretion rate in the underlying disc. The
advent of high resolution, spectro-imaging of jets by the {\it HST}
and ground-based observatories equipped with adaptive optics has
revealed their remarkable internal structure. Jets are strongly
correlated with the presence of protostellar accretion discs around
their host stars --- at least during the first million years or so of
their pre-main sequence evolution. Protostellar jet properties can be
directly probed by observations of the forbidden line emission that
they produce. Optical forbidden lines of oxygen, sulphur and nitrogen,
for example, are produced in very specific ranges of density and
temperature \citep[see][for a review]{Ray/etal:2007}. The
spectro-imaging results indicate that jets have an ``onion-like''
velocity structure, with the highest-speed outflow component, up to
400 km s$^{-1}$ in some sources, found at the core of the jet, while
the lower-velocity components are found systematically farther from
the outflow axis. The highest-resolution ({\it HST}) studies of the launch
region of TTS jets (about 50 AU) show gas densities $> 2 \times 10^4$
cm$^{-3}$, modest electron temperatures ($T \simeq 2 \times 10^4$ K),
and low ionization levels ($0.03$ -- $0.3$) \citep{Coffey/etal:2008}.
As we shall see, the onion-like velocity structure seen close to the
source is a reflection of the differential rotation of the underlying
Keplerian disc. Higher-resolution observations, down to the 1-AU scales
predicted for jet launch, would be highly desirable.

The discovery of jet rotation is perhaps one the most significant
advances of the last decade \citep{Bacciotti/etal:2002,
  Coffey/etal:2004}. By orienting the slit of the spectrograph
perpendicular to the jet axis at some distance along the jet, velocity
asymmetry that is compatible with jet rotation can be observed. Shifts
range from 5-25 km s$^{-1}$ at positions of 25 AU along the jet. From
the point of view of jet dynamics, these observations allow one to
measure the amount of angular momentum that is being carried by the
jet --- which ranges from $\sim 60$\% to $\sim 100$\% of the angular momentum
that would be carried through the disc by the observed disc accretion
rates. The ratio of mass carried in the jet to that moving through the
underlying accretion disc, $\dot M_{jet}/ \dot M_{a} \simeq
0.01-0.07$ \citep{Coffey/etal:2008}.

The strength of the magnetic field in the HH80-81 jet is estimated
from the synchrotron emission (on the basis of equipartition
arguments: see Section \ref{mjh:strength}), to be 0.2 milligauss
\citep{Carrasco-Gonzalez/etal:2010}. Even though the temperatures of
jets are quite low, the mechanism of diffusive shock acceleration has
been shown to be able to accelerate electrons moving at a few hundred
km s$^{-1}$ up to relativistic speeds. If this technique can be
exploited, we may indeed be on the threshold of being able to measure
all of the basic physical quantities that are needed to test the
physics of protostellar jets.

\subsection{Structure and dynamics of non-relativistic jets}

At the heart of the question of why jets are ubiquitous is the fact
that rapidly rotating magnetized bodies undergo strong spin-down
torques exerted by the threading fields. The basic theory of
magnetized winds from rapidly rotating stars was first worked out in
the seminal paper by \citet{Mestel:1968}. This theory focuses on
stationary solutions of axisymmetric flows from magnetized bodies ---
as applied in particular to rotating stars. It has turned out that
this theoretical framework is very useful in analyzing the structure
of jets and of the structure of their magnetic fields.

The overall dynamics of jets, and the physics of how they are launched,
is complicated. However, significant progress has been made by the
application of the theory of stationary, 2D (axisymmetric) MHD flows.
The usefulness of this highly simplified treatment arises from the
fact that we can derive conservation laws that turn out to be
important even for understanding the behaviour of stationary jets in
3D. First we decompose vector quantities into poloidal and toroidal
components (e.g.\ magnetic field $ {\bf B = B_p} + B_{\phi} {\bf \hat
  e_{\phi}} $). In axisymmetric conditions, the poloidal field ${\bf
  B_p}$ can be derived from a single scalar potential $a(r,z)$ whose
individual values, $a=\mbox{const}$, define the surfaces of constant
magnetic flux in the outflow and can be specified at the surface of
the disc \citep{Pelletier/Pudritz:1992}.

The joint conservation of mass and magnetic flux along a field line
can be combined into a single function
$k$ that is called the ``mass load'' of the wind,
which is a constant along a magnetic field line:
\begin{equation}
\rho {\bf v_p} = k {\bf B_p}.
\end{equation}
\noindent
This function represents the mass load per unit time,
per unit magnetic flux
of the wind. For axisymmetric flows, 
its value is preserved on each ring of  
field lines emanating from the accretion disc. 
Its value on each field line is determined
by physical conditions --- including dissipative
processes --- near the surface of the rotor.

The mass load plays  
a central role in jet dynamics (see Sections 3 and 4) and 
controls jet rotation,
collimation, and angular momentum extraction. 
It may be more revealingly written as
\begin{equation}
k(a) = { \rho v_p \over B_p} = {d \dot M_{\rm w} \over d \Psi},
\end{equation}
where $d\dot M_{\rm w}$ is the mass flow rate through an annulus of
cross-sectional area $dA$ through the wind and $d\Psi$ is the amount
of poloidal magnetic flux threading through this same annulus. The
mass load profile (as a function of the footpoint radius $r_0$ of the
wind on the disc) is determined by the physics of the underlying disc
and imposes an important boundary condition for all aspects of jet
physics.

The conservation of angular momentum along each
field line leads to the conserved angular momentum per unit mass;
\begin{equation}
l(a) = r v_{\phi}-{r B_{\phi} \over 4 \pi k} = \mbox{const}. 
\end{equation}
The form for $l$ reveals that the total angular momentum
is carried by both the rotating gas (first term) as well
by the twisted field (second term), the relative proportion
being determined by the mass load.

The value of $l(a)$ that is transported along each field line is fixed
by the position of the Alfv\'en point in the flow, where the poloidal
flow speed reaches the Alfv\'en speed for the first time ($m_{\rm
  A}=1$). Hence, the value of the specific angular momentum is found
to be
\begin{equation}
l(a) = \Omega_0 r_{\rm A}^2 = (r_{\rm A}/r_0)^2 l_0.
\end{equation}
where 
$l_0 = v_{K,0}r_0 = \Omega_0 r_0^2$ is the 
specific angular momentum of a Keplerian disc.  (Note, however,
that this treatment can easily be generalized to sub-Keplerian
discs.)
For a field line starting at a point $r_0$ on the
rotor (disc in our case), the Alfv\'en radius is
$r_{\rm A}(r_0)$ and this constitutes a lever arm for the flow.
The result shows that the angular momentum per unit
mass that is being extracted from the
disc by the outflow is a factor of $(r_{\rm A}/r_0)^2$ greater than
it is for gas in the disc.  For typical 
lever arms, one particle in the outflow can
carry the angular momentum of ten of its fellows left behind
in the disc.

The  conservation of energy along a field line
is expressed as a generalized version of Bernoulli's theorem.   
Since the terminal speed $v_p = v_{\infty}$ 
of the wind is much greater than its rotational speed, and
for cold flows, the pressure may also
be ignored, one finds the result: 
\begin{equation}
v_{\infty} \simeq 2^{1/2} \Omega_0 r_{\rm A} = (r_{\rm A}/r_0) v_{\rm esc,0}. 
\end{equation}

There are several important consequences for jet dynamics that follow
from this simple scaling. The first is that the terminal speed exceeds
the {\it local} escape speed from its launch point on an accretion
disc by the lever arm ratio. Furthermore, the scaling predicts that
the terminal speed scales with the Kepler speed as a function of
radius, so that the flow will have an onion-like layering of
velocities, with the largest velocities inside and the smallest on larger scales.
As we have already noted, such velocity structure is actually observed in the
dynamical structure of well-resolved jets. Finally, the terminal
speed depends on the depth of the local gravitational well at the
footpoint of the flow (the dependence on the local Keplerian rotation
speed in the case of a disc rotor) --- implying that it is essentially
scalable to flows from discs around YSOs of any mass and therefore
universal.

\subsection{Determining the jet launch region} 

A very useful combination of the energy and angular momentum conservation
that we derived above is given by the Jacobi constant along a field
line \citep[e.g.,][]{Pelletier/Pudritz:1992}; $j(a) \equiv e(a)-\Omega_0
l(a) $. This expression has been used \citep{Anderson/etal:2003} to
constrain the launch region on the underlying Keplerian disc.
Observers measure the jet rotation speed, $v_{\phi, \infty}$ at a
radius $r_{\infty}$, and the poloidal speed of the jet at this point,
$v_{p, \infty}$. Evaluating $j$ for a cold jet at infinity and noting
that its value (calculated at the foot point) is $j(a_0)=-(3/2) v_{\rm
  K,0}^2$, one can solve for the Kepler rotation at the point on the
disc where this flow was launched:
\begin{equation}
\Omega_0 \simeq v_{p,\infty}^2/ \left(2 v_{\phi,\infty} r_{\infty}\right). 
\end{equation}
When this relation is applied to the observed rotation of the 
Large Velocity Component (LVC) of the jet   
DG Tau \citep{Bacciotti/etal:2002}, this yields a range of disc radii 
for the observed rotating material 
in the range of disc radii, 0.3--4 AU, and the magnetic  
lever arm is $r_{\rm A}/r_0 \simeq 1.8$--$2.6$.   These results
show that an extended region of the disc can give rise to the 
jet, and that, moreover, the lever arm is in the predicted range
for the efficient launching of jets.  

\subsection{The disc/jet connection:  magnetic torques on discs}

The magnetized wind that is accelerated off the rotor carries away the
rotor's angular momentum. In the case of an accretion disc, the
angular momentum equation for the accretion disc undergoing a purely 
external magnetic torque (i.e., viscous torque neglected) may be written:
\begin{equation}
\dot M_{\rm a} { d (r_0 v_0) \over dr_0} = -r_0^2 B_{\phi} B_z \vert_{r_0, H},
\end{equation}
\noindent
where we have ignored transport by MRI turbulence or spiral waves.
Note that the toroidal field plays a fundamental role in disc structure
and accretion because it is so central to the action of a magnetic torque
upon the disc.  
By using the relation between the poloidal field and outflow on the 
one hand, as well as the link between the toroidal field and 
rotation of the disc on the other, the angular momentum equation 
for the disc yields one of the most profound scaling relations in disc wind
theory, namely, the link between disc accretion and mass outflow
rate:
\begin{equation}
\dot M_{\rm a} \simeq (r_{\rm A}/ r_0)^2 \dot M_{\rm w}.
\end{equation}
The observationally well-known result that, in many systems,
$\dot M_{\rm w} / \dot M_{\rm a} \simeq 0.1$ is a consequence of the 
fact that lever arms are often found in numerical and
theoretical work to be $r_{\rm A}/r_0 \simeq 3$ -- the observations
of DG Tau being a perfect example.
Finally, we note that the angular momentum that is observed to be carried by 
these rotating flows (e.g., in DG Tau) is a consistent fraction
of the excess disc angular momentum -- from 60--100\%
\citep{Bacciotti/etal:2004}, which is consistent with the high
extraction efficiency discussed here.

In reality, the evolution of discs is dictated by both viscous as well
as wind torques. The magnetic torque exerted on the underlying disc
has important effects upon its structure \citep{Ferreira:2002,
  Ferreira/Casse:2004}. Compared to standard $\alpha$-disk models,
\cite{Combet/Ferreira:2008} show that outflow torques modify the
underlying radial structure of discs, resulting in column density jumps
of a couple of orders of magnitude in going from the outer SAD
solution to the inner-jet dominated zone of the disc. The jet-dominated
region is also cooler and thinner.

\subsection{Jet collimation and toroidal fields}

In the standard picture of hydromagnetic winds, collimation of an
outflow occurs because of the increasing toroidal magnetic field in
the flow resulting from the inertia of the gas. Beyond the Alfv\'en
surface, the ratio of the toroidal field to the poloidal field in the
jet is of the order $B_{\phi} / B_p \simeq r/ r_{\rm A} \gg 1$, so
that the field becomes highly toroidal beyond the Alfv\'en surface.
Collimation is achieved by the tension force associated with the
toroidal field which leads to a radially inwards directed component of
the Lorentz force (or ``$z$-pinch''); $ F_{\rm Lorentz, r} \simeq j_z
B_{\phi}$.

The current carried by a jet can be easily written down:
\begin{equation}
 I = 
 2 \pi \int_0^r j_z(r',z')dr' = (c/2)
 r B_{\phi}. 
\end{equation}
The link between magnetic field structure and the collimation of jets
was made 
by \citet{Heyvaerts/Norman:1989}, where it was shown
that two types of solution are possible, depending upon
the asymptotic behaviour of the total current intensity in the jet.   In the limit that $I \rightarrow 0$ as
$r \rightarrow \infty $, the field lines are paraboloids
which fill space.  On the other hand, if the current
is finite in this limit, then the flow is collimated to cylinders.
The collimation of a jet therefore depends upon
its current distribution --- and hence on the radial distribution
of its toroidal field. 

It can be shown \citep{Pudritz/etal:2006} that, for a power-law
distribution of the magnetic field in the disc, $B_z(r_0, 0) \propto
r_0^{\mu-1}$, with an injection speed at the base of a (polytropic)
corona that scales as the Kepler speed, that the mass load takes the
form $k \propto r_0^{-(1+\mu)}$. In this regime, the current takes the
form $ I(r,z) \propto r_0^{-(\mu+1/2)}$. Thus, the current goes to
zero for models with $\mu < -1/2$, and these therefore must be wide
angle flows. For models with $\mu > -1/2$, however, the current
diverges, and the flow should collimate to cylinders. 

These results predict that jets should show different degrees of
collimation, depending on how they are mass loaded. As an example,
neither the highly centrally concentrated, magnetic field lines
associated with the initial split-monopole magnetic configuration used
in simulations by \cite{Romanova/etal:1997}, nor the similar field
structure invoked in the X-wind model \citep{Shu/etal:2000} should
become collimated in this picture. On the other hand, less centrally
(radially) concentrated magnetic configurations such as the potential
configuration of \cite{Ouyed/Pudritz:1997} and
\cite{BlandfordPayne1982} should collimate to cylinders.  By varying
both the density profile of the disc as well as its magnetic structure, \cite{Fendt:2006}
showed simulations in which highly collimated outflows from a ``flat'' radial
disk structure produced unsteady knots in the jet.  

The magnetic collimation picture, although it has considerable support
from theory and simulations, has been questioned by
\cite{Spruit2010}, who notes that the toroidal magnetic pressure in
jets will force them to expand. Collimation in this picture results
from magnetic fields and pressure in the external medium. Simulations
do indeed provide evidence for a regime in which magnetic bubbles
associated with simulations of protostellar outflows are observed to
expand radially. These cases arise in simulations of disc winds that
produce weak toroidal jet fields \citep{SeifriedII/etal:2011}. Strong
toroidal fields in these jets give rise to well collimated flows.

\subsection{Computer simulations: magnetic field structure of jets and their stability}

Numerical simulations have played an enormous role in exploring the
origin and evolution of jets. The pioneering stage of simulations of
jets from discs took place in the mid 1980s through the 1990s, with
the advent of magnetohydrodynamics (MHD) codes such as ZEUS. These early simulations recognized
that, since the mass load of the disc outflow is a key variable that
controls jet dynamics and structure, then the disc can be taken as a
boundary condition for the flow, with the initial magnetic field
structure and mass loading defined across this boundary. Simulations
demonstrated most aspects of the theory outlined above --- the
collimation of the jet, the concentration of jet density towards the
axis of the flow, the acceleration to high speeds, the ``onion-like''
velocity structure of the jet, and the dominance of the toroidal
magnetic field structure can be seen \citep[e.g.,][]{Ouyed/etal:1997,
  Ouyed/Pudritz:1997, Romanova/etal:1997, Krasnopolsky/etal:1999, Fendt:2006}.
Highly episodic behaviour of jets arises as a consequence of
sufficiently low mass loading of the jets \citep{Ouyed/etal:1997}.

\begin{figure}
\epsfxsize\textwidth
\epsfbox{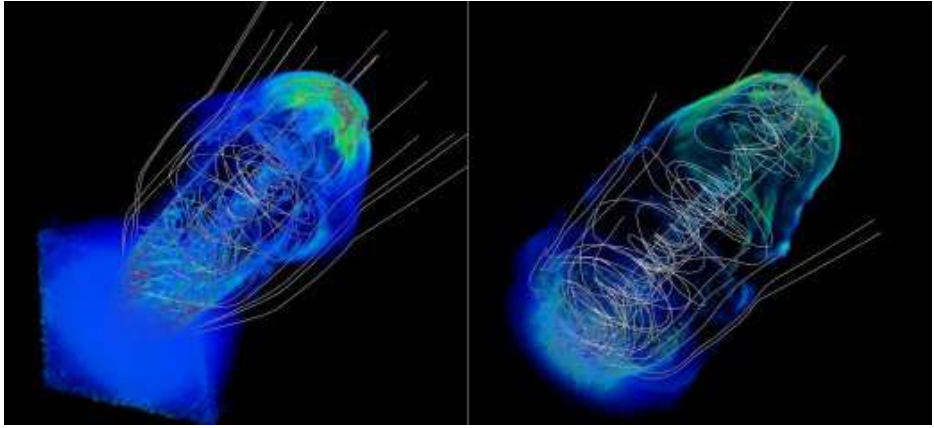}
\caption{Density and magnetic field lines for the Ouyed and Pudritz
  (1997) and Blandford and Payne (1982) models for the distribution of
  disc magnetic fields. The snapshots are shown as the jets have
  propagated out to 60 AU in the medium surrounding the magnetized
  disc. The protostar and the accretion disc are in the lower left
  corners of the panels, and hidden by the dense disc corona. Note
  that the core of the jet is dominated by poloidal field lines, while
  the outer regions are predominantly toroidal. Adapted from \cite{Staff/etal:2010}.}
\label{fig-maglines}
\end{figure}

Figure \ref{fig-maglines} shows ZEUS 3D simulations
of a jet being launched from an underlying disc
\citep{Staff/etal:2010}. The two panels show snapshots of the magnetic
field line geometry and jet density that arises within two particular
jet models --- those of \citet{Ouyed/Pudritz:1997} and
\citet{BlandfordPayne1982}. The field lines have a distinct toroidal
component that dominates the structure of the overall field on larger
scales. Poloidal magnetic field dominates towards the axis of the
outflow. The bow shock created by the jet is clearly seen in both
cases. The general structure of these models is rather similar. The
densest part of the jet are surrounded by a diffuse cavity-like region
that is dominated by the toroidal magnetic field. The more extended
outer material moves more slowly than the jet core.

If jets are dominated by toroidal fields far from the jet axis, why
are they stable? Theoretical studies of the linear stability of jets
show that they should become unstable when their speeds exceed the
local Alfv\'en velocity \citep{Ray:1981}. This was further
demonstrated by 3D simulations of idealized radio jets by
\citet{Hardee/Rosen:1999}. It turns out, however, that toroidally
dominated jets stabilize themselves through nonlinear processes.
Simulations show that jet instability, particularly in the form of
$m=1$ helical modes, builds up beyond the Alfv\'en
surface\citep{Ouyed/etal:2003}. These are stabilized by the magnetic
tension that stretches field in the linearly unstable region, which
ultimately increases the local Alfv\'en speed again. The net result is
a jet that is non-linearly stabilized by this mechanism, as well as by
the ``backbone'' provided by the poloidal field that dominates near
the outflow axis.

Since jets and accretion discs are highly coupled, jet torque and evolution will directly affect the disc.  Hence, there has been 
a major effort over the last decade to simulate jets and their discs as a single dynamical system \citep[e.g.,][]{Kudoh/etal:2002, Casse/Keppens:2002, vonRekowski/Brandenburg:2004}.   The important advantage of this approach is that one can avoid {\it ad hoc} descriptions of 
the mass loading and disc magnetic field line structure since these are self-consistently computed.    These simulations show that jet structure and dynamics retain the basic
features that the simpler simulations capture.   Moreover, simulations
by \cite{Zanni/etal:2007} which include the disk, and follow the
evolution of jets using the Adaptive Mesh Refinement code FLASH, find that more than 90\% of the gravitational potential energy liberated in the 
accretion flow is released into the jet.    Thus jets are highly
efficient in extracting both energy and angular momentum from the underlying
accreting systems.  

\subsection{Jets from magnetized  star-disc systems} 

So far we have emphasized disks as the underlying magnetized engines
for jets. However, collimated outflows may derive from both the
central stars as well as their surrounding disks. T-Tauri stars (TTS)
are well known to be significantly magnetized
\citep{Johns-Krull:2007}. As evidence that this may be occurring, we
focus on the fact that TTS are also known to rotate very slowly for
the amount of angular momentum that they are known to be accreting.
This has long been known to imply that there must be some external
torque being exerted upon the rotating star, carrying off or
intercepting the angular momentum. It has been proposed that this torque takes the form of an interaction of the star's magnetosphere and the
surrounding disk through closed magnetic field lines
\citep{Koenigl:1991}. Recent work suggests that most of the field in
such interaction is readily opened up by shear in the disk, leading to
very little spin-down torque and connection between star and disk
(e.g.. \cite{Matt/Pudritz:2004}. The interaction between inner disk
and stellar magnetosphere has also been hypothesized to give rise to a
so-called ``X-wind'', wherein the origin of jets and outflows is
pictured to be an MHD outflow --- based on the principles we discussed
above --- and originating from the very inner edge of the disk
\citep{Shu/etal:2000}. In this picture, dipolar stellar magnetic
field lines interacting with the inner edge of the disc are opened, leading
to  an outflow which 
intercepts angular momentum flowing inwards through the disc and
driving it out of the system in the outflow. One difficulty with this
models is in accounting for the huge amount of angular momentum that
jets are observed to transport --- 60\% or more of the angular
momentum transport through the disc at scales of 2 AU or so
\citep{Coffey/etal:2008}; another issue is the difference between
observed magnetic field strengths and the field strengths needed to drive the
X-winds \citep{Johns-Krull:2007}. A third class of models has been
proposed, wherein an accretion-powered wind from a magnetized, central
star carries off the disk angular momentum \citep{Matt/Pudritz:2005}.
The energy driving this wind is ultimately drawn from the
gravitational energy that is released by the infalling material from
the disc as it falls along magnetospheric field lines connecting the
star and the disc. A strong flux of Alfv\'en waves stimulated by the
impact of the accretion flow upon the star creates a wave-driven
wind \citep{Matt/Pudritz:2005, Matt/Pudritz:2008, Cranmer:2008}.

Simulations of such combined magnetized star-disk systems have been
carried out in which both the central rotating magnetized star and the
surrounding disk are treated as boundary conditions for the flows
\citep{Fendt:2009}. Compared to the case of pure disk winds, the
overall outflow is initially less collimated. This changes with time,
however, as a highly collimated flow emerges as a new dynamical state
across the grid when a stellar wind is included. The magnetic
interaction between the two wind components gives rise to reconnection
and large-scale flares as well, and the time-scale for the flaring is
long --- of the order of hundreds of inner disc rotation periods.
Finally, if the outflow is dominated by the stellar wind, the outflow
is either too weak or too high for low and high mass loads
respectively. The disk jet is required to have stable collimated flow
out to large scales.

The interaction of spinning, magnetized stars with their surrounding
disks have been extensively simulated for axisymmetric configurations
\citep{Romanova/etal:2002, Bessolaz/etal:2008}. Magnetospheric
accretion has been likened to ``funnel flow'' wherein matter falls onto
the star along field lines that are connected to the star and
co-rotate with it. The matter moves due to the gravitational force,
which dominates over the centrifugal force. The lifting of material
off of the disc and the onset of the funnel flow starts at the radius
at which the disk is slowed due to interaction with the more slowly
spinning magnetosphere. These simulations showed that the interaction
is time-dependent, with bursts of accretion followed by quiescent
periods in which material from the disc piles up against the
magnetosphere. Models usually assume that the disks are characterized
by a model ``$\alpha$-disc'' viscosity. In self-consistent 3D
simulations, accretion flow from a turbulent disk onto a tilted
magnetized star, where turbulence is driven by the MRI instability,
has been modelled \citep{Romanova/etal:2012}. The pattern of
alternating compression due to accretion and reconnection is also seen
in these 3D simulations. The magnetosphere truncates the disk at a few
stellar radii. The magnetic structure of the disk outside of the
stellar magnetosphere has a toroidal field $B_{\phi}$ that is a few
times larger than the poloidal field component. As we have noted
several times, the transport of angular momentum by twisted field
lines is an important part of the outflow mechanism. In the star-disc
interaction, the magnetic stresses dominate the matter stresses in the
disk and the pattern is quite inhomogeneus. This goes together well
with the predictions outlined in the basic theory above. Finally,
recent simulations of star-disc systems report the launch of fast
outflow at the interface in systems undergoing very high accretion
rates --- wherein the magnetic field in the star is strongly
compressed \citep{Lii/etal:2011}. No X-winds are observed for the much
lower accretion rates that dominate most of the star's accretion
\citep{Bessolaz/etal:2008}.

\subsection{Discs and jet formation during gravitational collapse}

We now leave the domain of stationary outflows, and examine how they
arise in the fully dynamical environments that characterize the birth of
stars. Protostellar jets appear during the earliest stages of star
formation and are the first obvious manifestation that gravitational
collapse and star formation are underway. No 3D simulations of purely
hydrodynamic collapse show that sustained jets can be produced in
collapsing systems. This may be why they were missed in early
hydrodynamical simulations of star formation. Add a threading magnetic
field, however, and the launch of energetic jets and outflows is
readily observed. The formation and evolution of jets takes place in a
few related phases: (i) the formation and early evolution of a
rotating, magnetized dense core, which is usually found embedded within
larger scale filaments within molecular clouds; (ii) the extraction of
angular momentum from the rotating core by torsional Alfv\'en waves
during the early formation and collapse stages, which significantly
de-spins the core; (iii) the gravitational collapse of a magnetized
core with the ensuing formation of a protostellar accretion disc; (iv)
the creation and launch of an outflow from the disc as it forms; and
(v) the accretion of the disc onto the central star and with it, the
gradual disappearance of the jet.

An important measure of the magnetization of the magnetized core is
given by the so-called mass-to-flux ratio, which is the ratio of the
gravitational to the magnetic energy of the system. Systems with supercritical
ratios (values greater than unity) cannot be
supported against collapse by their threading fields. The ratio can be
written as ~\citep{Mouschovias:1976}
\begin{equation}
 \mu = \frac{M_{\rm{core}}}{\Phi_{\rm{core}}}/\left(\frac{M}{\Phi}\right)_{\rm{crit}} = \frac{M_{\rm{core}}}{\int B_z dA}/\frac{0.13}{\sqrt{G}}\,.
 \label{eq:mu}
\end{equation}
A second parameter of importance is the ratio of the rotational to the
thermal energy of the core, known as $\beta_{rot}$. The phase space
spanned by these two parameters turns out to divide up the physical
properties of discs.

During the gravitational collapse of magnetized cores, the magnetic
field threading the region is wrapped up --- much like a twisting
rubber band. As this happens, an outward flow of torsional Alfv\'en
waves is created which transports angular momentum. Analytic solutions
of the braking torque that arises for idealized, disc-like rotors have
been written down. These have been found to fit the early stages of
simulations of the collapse of a magnetized, Bonner-Ebert sphere
\citep{Banerjee/Pudritz:2006}. As the collapse picks up, the winding
of the field becomes particularly strong at the accretion shock. It is
from this region that the first strong aspects of outflow begins. Soon
after this happens, an inner region of outflow is launched which is
centrifugally driven.

For fields that are sufficiently supercritical, the magnetic braking
can be so significant that the disc material is left rotating at
significantly sub-Keplerian levels. This has raised the question of
whether or not discs can even form in the reasonably magnetized
environments that are found in regions of star formation
\citep{Hennebelle/Fromang:2008, Mellon/Li:2008}. The results depend
quite sensitively on the initial conditions chosen --- i.e., the
mass-to-flux ratio. At values of $\mu \simeq 2$ -- 10 , the resulting
systems are strongly affected. This intriguing result points to the
importance of having appropriate astrophysical initial conditions.
Recent simulations that feature long-term evolution of discs using
``sink'' particles find that in fact a reasonably Keplerian initial
disc can form during the long-term evolution of the collapse (Duffin
et al., in preparation). A grid of models for the formation of massive
stars shows that reasonably quickly rotating cores and lower
magnetized cores produce Keplerian discs \citep{SeifriedI/etal:2011}.

\begin{figure}
\epsfxsize\textwidth
\epsfbox{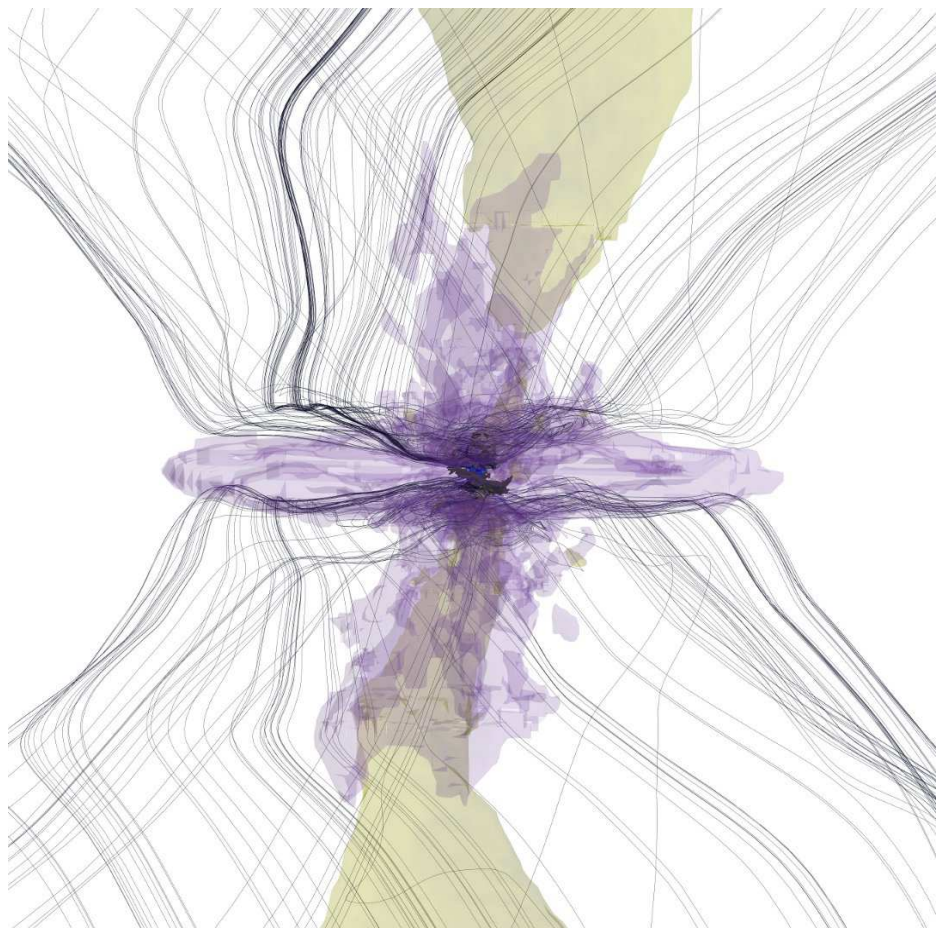}
\caption{Magnetic field structure in a collapsing magnetized core. A
  flattened structure is produced by the gravitational collapse.
  The purple contours correspond to $3.3\times10^{-17}$ g cm$^{-3}$
  ($n=8.5\times10^{6}$ cm$^{-3}$) and black contours to
  $1.3\times10^{-15}$ g cm$^{-3}$ ($n=3.4\times10^{8}$ cm$^{-3}$).
  Yellow contours correspond to outflowing velocities $v_z > 1.5$ km
  s$^{-1}$, and extend to a height of $10^4$ AU.
  Black lines are
  magnetic field lines. Adapted from Duffin et al.~(2012). }
\label{fig-collapse}
\end{figure}

Figure \ref{fig-collapse} shows a zoomed-in picture of the magnetic
structure that arises in the collapse of a solar mass system of a low
mass star formation (Duffin et al.\ 2012, in preparation). The jet
produced in this simulation reaches out to the extent of the initial
Bonner-Ebert sphere, of about $10^4$ AU in diameter. The gravitational
collapse has pulled in the field lines which thread the disc. A
transient, large disc of 2000 AU in extent has formed, which
ultimately fragments into a ring and an inner 100 AU scale disc within
an outer disc rotation period. The innermost region of the disc is
seen to warp, which is a consequence of the torques induced by the
magnetized wind. The outflow, as a consequence, precesses with the warp
in the disc. Precessing jets are common and the simulation represented
in this image shows, for the first time, that magnetic fields can
cause this behaviour.
 
The launch mechanism of the outflows during collapse is under some
current debate. For low-mass star formation ($M_{\rm{core}} \sim 1
M_{\odot}$), the generation, evolution and properties of protostellar
outflows have been studied in great detail over the last
years~\citep[e.g.,][]{Banerjee/Pudritz:2006, Mellon/Li:2008,
  Hennebelle/Fromang:2008, Duffin/Pudritz:2009}. The two basic
mechanisms that have been discussed in the literature focus on whether
the outflow is driven by centrifugal
acceleration~\citep{BlandfordPayne1982,Pudritz/Norman:1986,Pelletier/Pudritz:1992,
  Ferreira:1997} or by the pressure of the toroidal magnetic field
that was wrapped up in the initial collapse
~\citep{Lynden-Bell:1996,Lynden-Bell:2003}. The second mechanism ---
known as a tower flow --- is highly transient and does not lead to
sustained outflow moving out from the source. It is based on the study
of equilibria of highly wound magnetic structures. General energy
theorems demonstrate that they form tall magnetic towers, the height
of which grows with every turn at a velocity related to the circular
velocity in the accretion disc. The pinch effect amplifies the
magnetic pressures toward the axis of the towers.

A more generalized criterion for outflow in disc wind theory, assuming
that the gas is not strictly co-rotating with the field, has recently
been proposed \citep{SeifriedII/etal:2011}. This model shows that
disc-wind theories can have acceleration mechanisms that are driven by
centrifugal forces as well as toroidal fields. Both magnetic tower and
disc-wind theories can be commonly understood as a consequence of the
same disc-wind models. It is of particular interest that the MHD flow
equations discussed above actually have two aspects to them. Close to
the outflow axis, it may be shown that the toroidal field generation
is not significant, and there the outflow is driven like a centrifuge.
Farther away, however, a toroidal field component is inescapably
produced. It can be shown that the pressure gradient associated with
this contributes to pushing a continuous outflow from the outer
regions of the disc.

\begin{figure}
\epsfxsize0.5\textwidth
\epsfbox{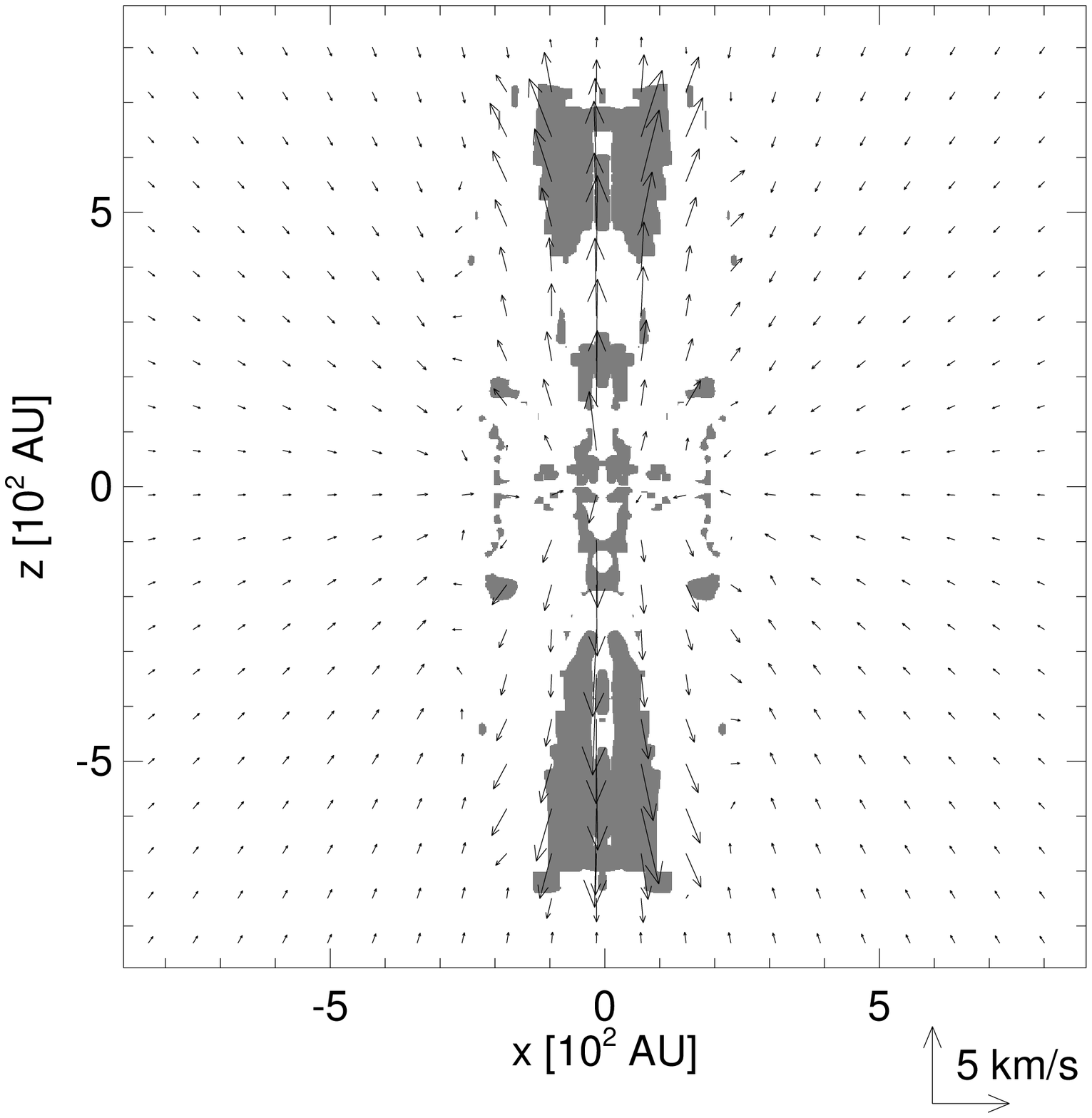}
\epsfxsize0.5\textwidth
\epsfbox{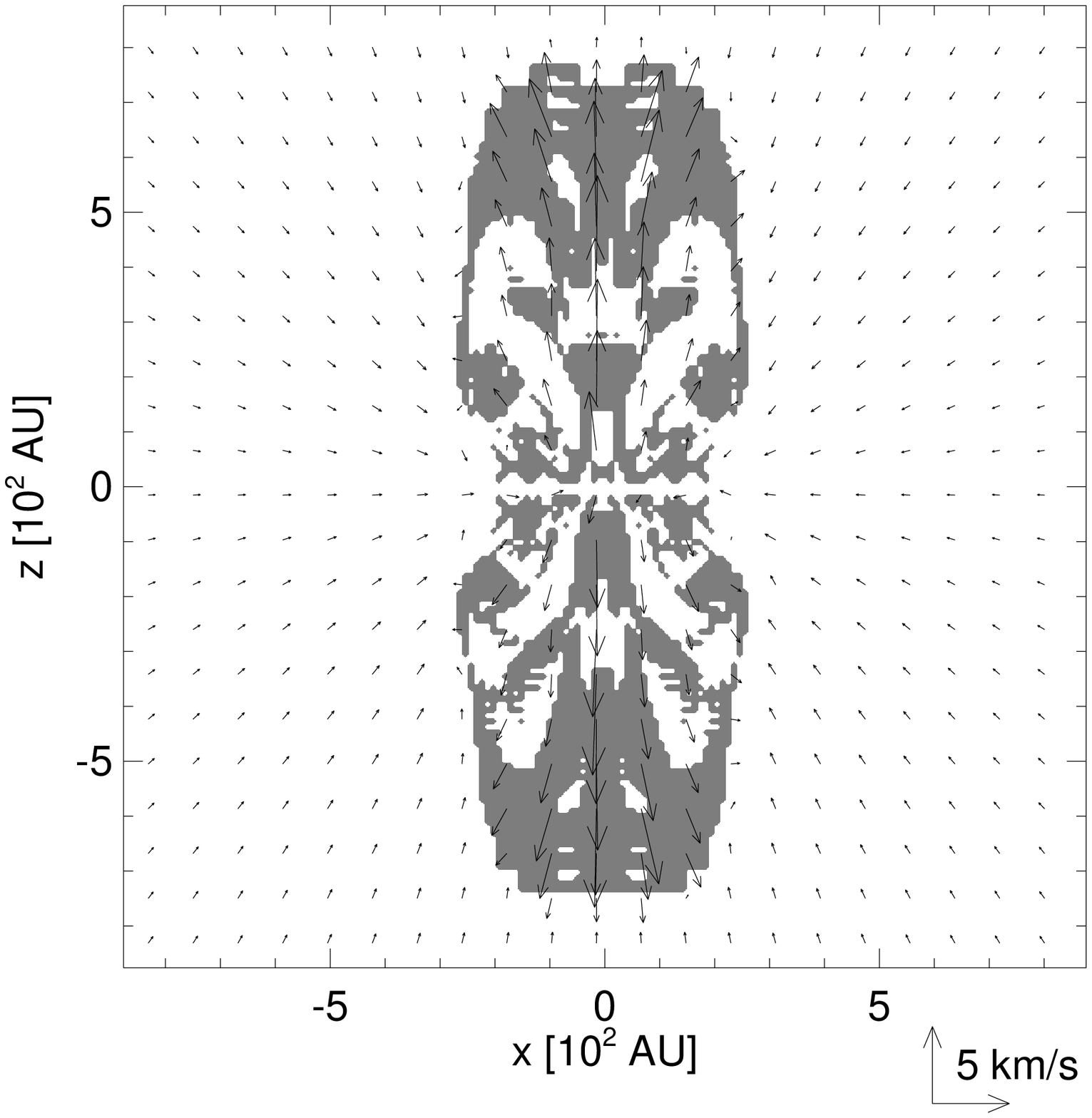}
\caption{Left: the region of jet that is driven by centrifugal acceleration --- denoted in grey --- in a simulation of a 
collapsing, solar mass, magnetized Bonner-Ebert sphere. Right: as
left, but grey regions show the regions controlled by the full
acceleration mechanism, including toroidal pressure gradients.
Adapted from Duffin et al.\ (2012).
}
\label{fig-duffin}
\end{figure}

The two regions of the jet may be seen in Fig.\ \ref{fig-duffin} where
we compare the size of the region controlled by purely centrifugal effects
with those arising from toroidal pressure gradients. This illustrates
that magnetic fields in jets play different roles depending upon where
they are found. The difference between these two snapshots is the
region in the jet where gas is driven primarily by toroidal field
gradients. Note that these dominate in the outer parts of the jet,
whereas the centrifugal driving predominates in the region close to
the jet axis.

\subsection{Relativistic considerations and relation to protostellar jets}

There is nothing in the physics of jets that suggests that
relativistic sources, such as microquasars and AGN, and should have
radically different driving mechanisms from those of protostellar
systems --- indeed, the dominant magnetic models make use of similar
MHD equations. The fact that both a sufficiently rapidly rotating
black hole and/or the surrounding magnetized disc are sources for
relativistic jets has an obvious parallel with the protostellar
systems. The theory and simulations of non-relativistic jets, as
applied to protostellar systems, show some remarkable connections with
the observations of AGN jets, which are discussed in Section 3 and
Section 4. There are some important differences, however, which we
discuss here.

The theory of relativistic jets shows that the essential parameter is the {\em magnetization}
parameter \citep{Michel:1969, Camenzind:1986},
\begin{equation}
\sigma = \frac{\Psi^2\Omega_{\rm F}^2}{4 \dot{M} c^3}.
\label{eq_sigdef}
\end{equation}
where the  iso-rotation parameter $\Omega_{\rm F}$ is  the
angular velocity of the magnetic field lines.
The function $\Psi = B_p r^2$ is a measure of the magnetic field distribution and $\dot{M}\equiv \pi\rho v_p R^2$
is the mass flow rate within the flux surface.
Equation (\ref{eq_sigdef}) demonstrates that the launch of a highly relativistic
jet requires at least one of three conditions
 ---  rapid rotation, a strong magnetic field, and/or a
comparatively low mass load.   A strong magnetic field lowers the 
mass load in the jet, with the consequence that more Poynting flux can
be converted into kinetic
energy per unit mass flux.

In the case of a spherical outflow ($\Psi$ = constant) with negligible gas
pressure one may derive the Michel scaling  between the asymptotic Lorentz
factor and the flow magnetization \citep{Michel:1969},
\begin{equation}
\Gamma_{\infty} =
\sigma^{1/3}
\label{eq_sigmich}
\end{equation}
Depending on the exact magnetic field distribution $\Psi(r,z)$,
in a {\em collimating jet} the matter can be substantially accelerated
beyond the fast magnetosonic point \citep{Begelman/Li:1994, Fendt/Camenzind:1996}.
As a result, the power law index in eq.\ (\ref{eq_sigmich}) can be
different from the Michel-scaling
 \citep{Fendt/Camenzind:1996, Vlahakis/Konigl:2003}.

An essential difference between relativistic and non-relativistic flows is the role of electric fields.
They are negligible in non-relativistic flows but are comparable to magnetic fields
in the relativistic case.   The issue is that these fields can lead to the decollimation of the jet.  Current-carrying relativistic jets have, however, been demonstrated to collimate \citep{Chiueh/etal:1991}.  
Relativistic jets have another feature with 
no counterpart in non-relativistic systems, namely the {\em light
  cylinder},  located at the cylindrical radius
$r_{\rm l}  = c/\Omega_{\rm F}$.
At the  light cylinder the velocity of the
magnetic field lines ``rotating'' with angular velocity $\Omega_{\rm F}$
coincides with the speed of light.  Outside the light cylinder, the magnetic field lines ``rotate'' faster than the speed of
light. As the field line is not a physical object, the laws of physics
are not violated. 
The light cylinder is interpreted as the Alfv\'en surface in the limit
of vanishing matter density (force-free limit).
Crucially, {\it the location of the light cylinder determines the relativistic character of the
magnetosphere. If the light cylinder is comparable to the dimensions of the object
investigated, a relativistic treatment of MHD is required}.  

Simulations of relativistic axisymmetric MHD disc winds have demonstrated that jets do indeed collimate,  with 
half opening angles of 3--7$^\circ$ \citep{Porth/Fendt:2010}.   These authors set up a relativistic analogue
of the \cite{Ouyed/Pudritz:1997} simulations, and emphasized flows far enough
away from the central black hole that general relativistic effects could be ignored.  The outflow
has a mixed character --- a combination of the Blandford-Payne, centrifugally driven wind with
a toroidal pressure-dominated flow of the type discussed by \cite{Lynden-Bell:1996}.   Lorentz factors
of the order $\Gamma = 6$  are produced.  The outflow has an interesting substructure, with a 
narrow relativistic jet towards the axis that is surrounded by a sub-relativistic flow launched from 
further out in the disc.   This structure agrees with the observations of AGNs on the pc scales, as
we discuss in Section 3.  Finally we note that the toroidal fields that are generated in 3D simulations 
of jets seem to act as a sheath that protects the spine of the jet core from losing momentum to the surrounding
medium \citep{Migone/etal:2010}.    
   
The advent of 3D general relativistic, MHD (GRMHD) codes (eg.
\cite{McKinney:2006, McKinney/etal:2012}) has opened up the study of the
role of black holes in the origin and evolution of AGN jets. As for
the purely relativistic case, one of the primary questions is to determine
how relativistic the jets become, and to understand what maintains
their stability and collimation. It is generally thought that there
are two regimes for AGN jets: a magnetized disk surrounding a black
hole, as described by the \cite{BlandfordPayne1982} picture, and/or
the spinning black hole whose ergosphere is threaded by magnetic
field, as originally envisaged by \cite{Blandford/Znajek:1977}. Most
theories and the simulations suggest that it is the spinning black hole that
produces the highly relativistic jets, while lower-speed outflows
arise from the surrounding disk. Recent GRMHD simulations
\citep{McKinney/Blandford:2009} model the initial threading magnetic
field geometries for the spinning black hole as either a dipole or
quadrupole configuration, while a thick torus is used to model an AGN
disk. In addition to the need for significant spin of the central hole
(the spin parameter of the hole is $a/M > 0.4$), their results
demonstrate that the initial magnetic configuration also plays a
significant role. Jets produced by an initial dipole structure can
achieve high Lorentz factors, ($\Gamma > 10$) and remain highly
collimated with an opening half-angle of $\theta_j \simeq 5^o$ at
$10^3$ gravitational radii. This is a stable jet with the substructure
dominated by a stable $m=1$ mode, probably excited in the underlying
turbulent torus.

However, this does not occur for a more complex, quadrupolar, initial
magnetic geometry, where the jet does not reach steady realistic
speeds ($\Gamma < 3$). The quadrupolar field leads to mass loading
which considerably slows the jet, and makes it prone to disruption.
This outflow has a weak and disorganized poloidal field and a much
stronger (by factor of 10-40) toroidal field. The authors speculate
that these differences with the dipole jet may have something to do
with the observed morphology of extragalactic jets; in particular,
their classification into weakly relativistic ``FR I-type'' jets
associated with AGNs in more clustered regions, and highly
relativistic ``FR II-type'' jets associated with isolated objects. The
latter would be associated with the spinning black holes and the
former with the underlying accretion disks.

There are many similarities between these relativistic results and the
non-relativistic work and theory described earlier. The stability of
jets over a large variety of scales seems assured for both regimes ---
they are able to run with stable substructure, dominated by an $m=1 $
mode, without falling apart. Both types of systems can remain very
highly collimated, and this is related to their mass loading which is
connected with their magnetic geometry. Moreover, toroidal magnetic
field dominates the non-relativistic jet structure, which again is a
likely consequence of the theory we have outlined in this Section.

In AGN systems, there is a concerted effort to understand jets on the
smallest scales, which correspond to a few pc, in order to test
theoretical models. The analysis described in this Section has shown
that magnetic fields in jets are poloidal near the jet axis, and
dominated by toroidal field farther out. This is a natural consequence
of the acceleration and collimation mechanisms that should also apply
to non-relativistic (eg. protostellar) and relativistic (AGN)
accretion systems. We are now just beginning to be able to simulate
the evolution of compact jets to larger scales (several pc) which
allows direct comparison with observation. Simulations indicate that
the field in protostellar jets becomes highly wrapped near the
bow-shock of the jet \citep{Staff/etal:2010}, and this may have
consequences for understanding AGN jets as well. In the next Section,
we turn to the parsec-scale structure of AGN jets, to connect
observations with the theory and simulations presented above.

\section{Structure and dynamics of parsec-scale AGN jets}

\subsection{Backdrop for current studies}
\label{dcg:backdrop}

Radio studies of jets from Active Galactic Nuclei (AGNs) have
primarily focused on parsec scales (with centimetre-wavelength Very
Long Baseline Interferometry, VLBI, giving milliarcsecond resolution)
and kiloparsec (kpc) scales (with arcsecond-resolution interferometers
such as the Very Large Array and MERLIN), with relatively few
observations probing intermediate scales. This means that, although
the pc- and kpc-scale observations in some ways tell a coherent and
self-consistent story, in others they seem to show discrepancies, with
the connection between the behaviours observed on the two scales not
always being clear. One reason for this may be that, although the
parsec- and kpc-scale jets form different parts of a single structure,
the observations may be determined by different contributions from
``global'' properties intrinsically related to the jet and ``local''
properties due to localized perturbations, interactions with the
surrounding medium, variations in the magnetic field and density of
the ambient medium, etc.

Throughout most of the period in which AGN jets have been studied with
VLBI, since the early 1980's, interpretations of observed features and behaviour
have focused on the possibility that the observed jet properties are
associated primarily with local agents, in particular, relativistic
shocks propagating in the jets. From a certain point of view, this was very
natural, and was motivated by factors such as the visibly inhomogeneous
appearance of the jets themselves (they are often dominated by distinct
components) and the common occurrence of variability. A theoretical framework arose 
in which the jet components were primarily shocks propagating along the jet,
with various processes related to these shocks giving rise to the observed
very rapid variability. This picture seemed to receive support from the 
earliest VLBI polarization observations, which showed that an appreciable
number of AGN jet components have magnetic fields that are predominantly
transverse to their jets --- this was interpreted as reflecting the compression
of an initially tangled magnetic field by a transverse shock, causing the 
field to become aligned in the plane of compression~\citep{Gabuzda1992}. 

More recently, various theoretical and observational studies have begun to 
explore the possibility that at least some of the polarization structures 
detected on parsec scales are associated with the intrinsic magnetic fields 
of the jets themselves. One area of such research involves theoretical 
simulations and observational searches for evidence of helical or toroidal 
magnetic fields carried by the jets, which are expected to form as aresult 
of the joint action of the rotation of the central black hole and its 
accretion disc and the jet outflow. The presence of toroidal or helical
magnetic fields should give rise to characteristic transverse intensity and linear
polarization structures across the jets, as well as transverse gradients in 
the observed Faraday rotation, due to the systematic change in the 
line-of-sight magnetic field across the jet. 

Thus, one of the challenges currently faced by researchers studying AGN jets
is to try to discern which features are primarily due to the action of  local 
agents, such as shocks or interactions with the ambient medium, and which are
primarily associated with the intrinsic magnetic field and other properties 
of the jets themselves.  This Section will review recent observational results 
for high-resolution (VLBI) observations of AGN jets, together with some
related theoretical simulations, in the framework of this challenge. 
 
\subsection{Estimates of VLBI-core $B$-field strengths}
\label{dcg:strength}

The direction of the magnetic field {\bf B} giving rise to synchrotron radiation 
can be deduced from the direction of the associated observed electric vector
position angle (EVPA), if the optical depth regime is known: in the absence of 
Faraday rotation,
{\bf B} is perpendicular to the EVPA in optically thin regions and parallel to
the EVPA in sufficiently optically thick regions \citep[see, e.g.,][]{GabuzdaGomez2001}. Thus, it is relatively
straightforward to get an idea of the overall magnetic-field configuration even
from VLBI observations at a single frequency, as long as the effects of Faraday
rotation are not too severe.  However, it is not possible to estimate the 
strength of the synchrotron $B$ field directly using observations at a single
frequency: multi-frequency observations are required, and even then, somewhat
indirect techniques must be applied. 

\begin{figure}
\epsfxsize\textwidth
\epsfbox{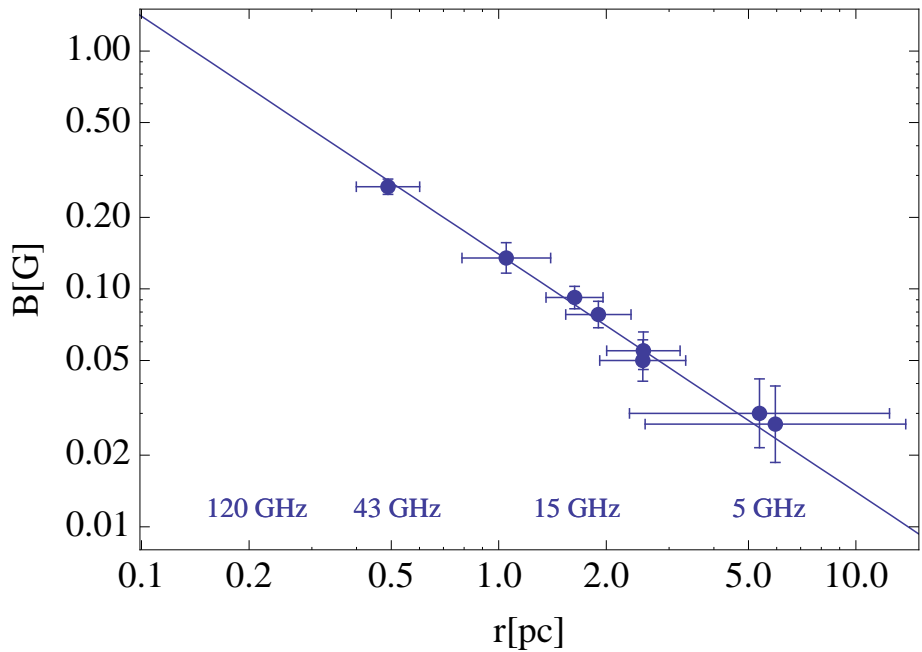}
\caption{Example of a derived dependence of $B$-field strength on distance from
the jet base, approximately $B\propto r^{-1}$ \citep{OSullivanGabuzda2009B}.}
\label{fig-parsec-bfield}
\end{figure}

One of these is based on the frequency dependence of the position of
the observed VLBI core in a Blandford--K\"onigl jet
\citep{BlandfordKonigl1979}. In this picture, the VLBI core is
essentially like a photosphere, and corresponds to the surface where
the optical depth $\tau$ is equal to unity. The position of this
$\tau=1$ surface is located further down the jet at lower frequencies:
$r\propto\nu^{1/k_r}$, where $r$ is distance from the jet base, $\nu$
is frequency and $k_r$ is a parameter that, in general, depends on the
spectral index and the laws for the decline of the electron density
and magnetic field with distance from the central engine
\citep{Konigl1981}. In practice, the observed VLBI core includes
emission from both the vicinity of this $\tau=1$ surface and the
innermost jet, but the essence of this basic theoretical picture
holds. In contrast to the behaviour of the core region, optically thin
regions --- in other words, regions in the jet --- are expected to
coincide at different frequencies. By aligning optically thin regions
observed simultaneously at different frequencies and deriving the
relative positions of the observed VLBI cores, the resulting ``core
shifts'' can be used to derive estimates of the core magnetic fields,
if reasonable estimates for the Doppler factor, viewing angle and jet
opening angle are available \citep{Lobanov1998}. This technique has
only been applied to a handful of objects, but results obtained so far
seem to give similar results for the cores of different AGNs,
indicating that $k_r\approx 1$ in most cases (which corresponds to
conditions not far from equipartition) and typical $B$ fields at a
distance of 1~pc from the jet base of about 0.15~Gauss
\citep{OSullivanGabuzda2009B}. Although these estimates are somewhat
uncertain, due to uncertainties in the estimates of the Doppler factor
and viewing angle, which cannot be derived directly from observations,
the fact that similar core-region $B$ fields are derived for various
different AGNs suggests that these estimates are reasonably correct,
at least to order of magnitude. These same studies have yielded
evidence that the magnetic field falls off with distance from the jet
base roughly as $r^{-1}$ (Fig.~\ref{fig-parsec-bfield}).

\subsection{Basic observational properties of the polarization of AGN jets}
\label{dcg:properties}

Although a wealth of polarization structure is observed among the variety of
AGNs that have been studied with VLBI, it is possible to identify certain basic
characteristics of the jet polarization. The most important of these is that,
in a substantial majority of cases, the jet polarization is oriented close to
parallel or perpendicular to the local jet direction (e.g.,
\citealt{ListerHoman2005}). The effect of Faraday
rotation is usually modest outside the core regions, leading to polarization
rotations of no more than $10-15^{\circ}$ at centimetre wavelengths (see also
Section~\ref{dcg:helical}). For this
reason, this tendency for the jet polarization to be either parallel or 
perpendicular to the local jet direction was evident even in the earliest
VLBI polarization observations carried out at a single wavelength (e.g.,
\citealt{Gabuzda1992,Cawthorne1993}).  On average, there is a
tendency for AGNs with relatively luminous optical line emission (``quasars'')
to display higher apparent superluminal speeds and to most often have 
primarily longitudinal jet magnetic fields, while those having relatively 
weak optical line emission (``BL Lac objects'') generally display lower 
apparent jet speeds and most often have predominantly transverse jet 
magnetic fields (e.g., \citealt{Gabuzda1994,Britzen2008}).

Jet regions with oblique, or ``misaligned'', polarization (i.e., neither 
close to parallel or perpendicular to the 
local jet direction) that cannot be ascribed to Faraday rotation are sometimes
observed, but are comparatively rare. Jets with extensive regions of
longitudinal or transverse $B$ field are observed. Little is known about
the nature of ``inter-knot'' polarized emission located between bright jet
features; VSOP space-VLBI polarization observations have revealed some cases
in which longitudinal inter-knot polarization, implying transverse $B$ fields 
associated with
this underlying jet emission, are observed \citep{Gabuzda1999,Pushkarev2005}.

\begin{figure}
\epsfxsize\textwidth
\epsfbox{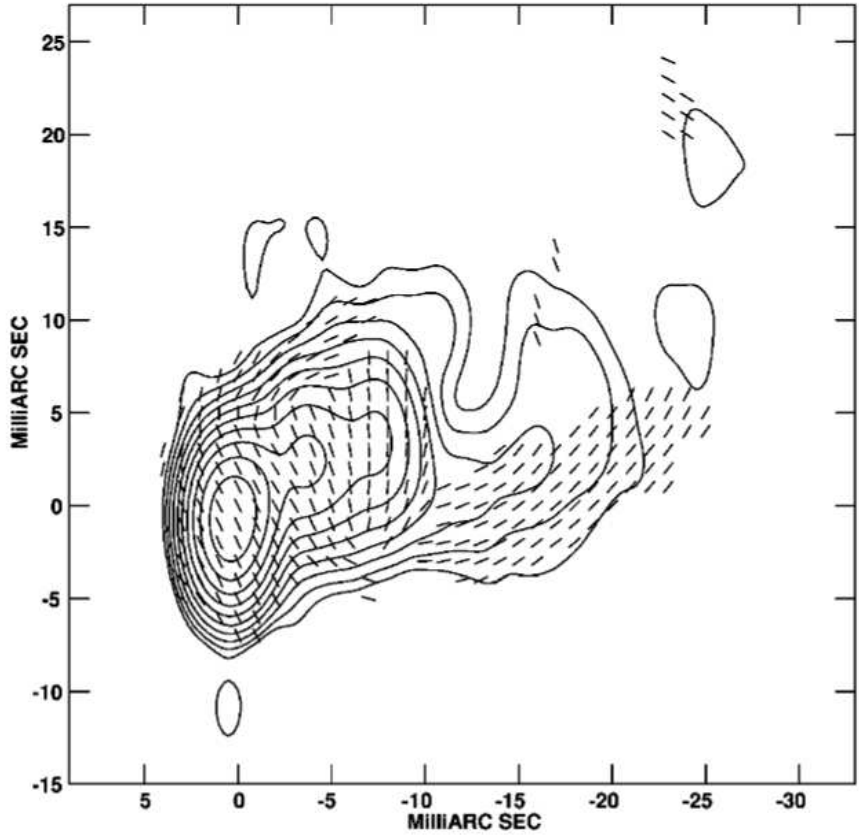}
\caption{Example of ``spine+sheath'' polarization structure across an
  the jet of the blazar 1055+018 \citep{Attridge1999}. The object is
  at a redshift $z = 0.888$, so 1 mas equals 7.78 pc. The vectors show the direction of the $B$-field.}
\label{fig-1055-bfield}
\end{figure}

Appreciable transverse polarization structure is observed in many jets, as
becomes clear through a visual inspection of 15-GHz maps produced by the
MOJAVE VLBA monitoring project (e.g., \citealt{ListerHoman2005}). In some 
cases, ``spine+sheath'' polarization
structures are observed, with longitudinal
polarization (orthogonal $B$ field) near the central axis of the jet and 
orthogonal polarization (longitudinal $B$ field) near the jet edges (e.g.
\citealt{Attridge1999,Pushkarev2005}; Fig.~\ref{fig-1055-bfield}). In other
cases, the predominant polarization is orthogonal to the jet, and is offset
appreciably from the central axis of the jet.

\subsection{Interpretation of observations}
\label{dcg:interpretation}

Jet knots whose dominant polarization is longitudinal, implying a transverse
$B$ field, have frequently been interpreted as transverse shocks that have
compressed an initially tangled magnetic field so that it has become ordered
in the plane of compression \citep{Laing1980,HughesAllerAller1989}, 
while the presence of orthogonal
polarization (longitudinal $B$ field) has been ascribed to the action of shear
that has enhanced this component of the field in a layer surrounding the jet.
Transverse polarization structures have similarly sometimes been interpreted as
indicating the combined action of shocks and shear (e.g., 
\citealt{Attridge1999}).
In the framework of the empirical connection between higher/lower apparent jet 
speeds and predominantly longitudinal/transverse jet magnetic fields, it has
been suggested that transverse shocks may form more easily in jet outflows 
that have lower intrinsic speeds \citep{Gabuzda1994,DuncanHughes1994}. 

Alternately, all of these polarization structures can be understood as possible
manifestations of helical magnetic fields carried by the jets (e.g., 
\citealt{LyutikovParievGabuzda2005}): for example, jets with predominantly orthogonal 
or longitudinal 
$B$ fields could carry helical fields with relatively large or small pitch angles. 
The tendency for AGNs displaying higher/lower apparent superluminal speeds 
to be more likely to have primarily longitudinal/transverse jet magnetic fields 
is interesting in this connection. One possibility is that 
this could reflect a relationship between the
pitch angle of a helical jet $B$ field and the physical speed of the jet outflow:
the ratio of the speed of the jet outflow to the rotational speed of the central 
black hole and accretion disc could be higher for quasars, leading to smaller
pitch angles for the helical $B$ fields threading their jets and a dominance of
the longitudinal component of the helical field, whereas the low outflow speeds
of BL Lac objects lead to higher pitch angles for their helical fields and a
dominance of the toroidal component of the helical field.

The appearance of a ``spine+sheath'' transverse polarization structure or
polarization, implying a longitudinal $B$ field offset toward from the central
axis of the jet, could likewise come about if the jet
carries a helical magnetic field, with the azimuthal component dominating near
the central axis of the jet and the longitudinal component becoming dominant
near the edges.  The different viewpoints advocated by these two different
types of interpretation essentially reflect the challenge referred to above,
of distinguishing between characteristics associated with the action of local
agents and those associated with intrinsic properties of the jets themselves.

Considerable progress has also been made recently in using the results of theoretical 
computations and simulations of AGN jets to generate simulated VLBI intensity, 
spectral-index, polarization and Faraday-rotation distributions that can, in 
principle, be compared with observations. Examples include the studies carried out 
by \cite{Zakamska2008}, \cite{Gracia2009}, \cite{Mimica2009}, 
\cite{BroderickMcKinney2010}, \cite{Porth/Fendt:2010}, \cite{Porth2011} and 
\cite{Clausen-Brown2011}.  \cite{Mimica2009} and \cite{Mimica2010, Mimica2012} 
have considered various observational signatures of the presence of shocks 
and shock-heated gas in relativistic jets, including jets carrying helical 
fields.  \cite{BroderickMcKinney2010} and \cite{Porth2011} have generated 
theoretical Faraday-rotation images based on computations and simulations of  
AGN jets with a helical magnetic-field component. \cite{Porth2011} also 
considered the behaviour
of the polarization angle as a function of the wavelength $\lambda$ squared in the 
(partially) optically thick core region, demonstrating deviations from a 
$\lambda^2$ relation away from a limited range of relatively short wavelengths;
this effect appears to have been observed by \cite{OSullivanGabuzda2009RM}.  
One limitation here is the need to use some
scheme to extrapolate results for small scales to the scales accessible to 
observations with VLBI. However, such studies are beginning to enable much
closer comparisons with observations than was possible previously, and are being
used to identify a range of characteristic observational signatures of various
types of jets and jet magnetic fields. It is clear that the ability to reliably
measure transverse structure in intensity, spectral index, polarization and 
Faraday rotation across AGN jets will be key to probing the  nature of the jet
magnetic fields.

\subsection{Observational evidence for helical or toroidal magnetic fields}
\label{dcg:helical}

As is discussed in Section~2 above, helical jet magnetic fields are predicted 
by a wide range of theoretical models 
and numerical simulations (e.g. \citealt{BlandfordPayne1982}, 
\citealt{Ouyed/Pudritz:1997}, \citealt{Krasnopolsky/etal:1999}, 
\citealt{Fendt:2006}), making observational searches for evidence for helical 
magnetic fields associated with AGN jets of considerable interest. The
most promising approach in this area is searching for transverse
gradients in the Faraday rotation measure distribution across AGN
jets, as was first pointed out by \cite{Blandford1993}. The observed
Faraday rotation is proportional to the integral along the line of
sight of $n_e {\textbf B}\cdot d{\textbf l}$, where $n_e$ and {\textbf
  B} are the electron density and magnetic field in the region of
Faraday rotation and the length element $d${\bf l} points along the
line of sight toward the observer. Therefore, the magnitude of the
Faraday rotation depends on the line-of-sight component of the ambient
magnetic field, and the sign of the Faraday rotation is determined by
the direction of this component of the field. Therefore, if a jet and
its immediate vicinity are threaded by a helical magnetic field, this
field will give rise to a systematic transverse gradient in the
observed Faraday rotation, due to the systematic change in the
line-of-sight component of the helical $B$ field across the jet. The
detection of such transverse Faraday-rotation gradients is
observationally challenging, since it involves a joint analysis of
multi-frequency radio images, and the observing conditions are not
optimal on either parsec or kpc scales: on parsec scales, the jets are
very compact, so that they are sometimes only marginally resolved in
regions of strong polarization, even when observed with VLBI, while,
on kpc scales, the jets are better resolved, but the effect is weaker
due to the increased distance from the base of the jet. To obtain
increased sensitivity to Faraday rotation, one can observe at longer
wavelengths, however, this comes at the expense of a corresponding
loss of resolution.

\begin{figure}
\epsfxsize\textwidth
\hskip -1in\rotatebox{-90}{\epsfbox{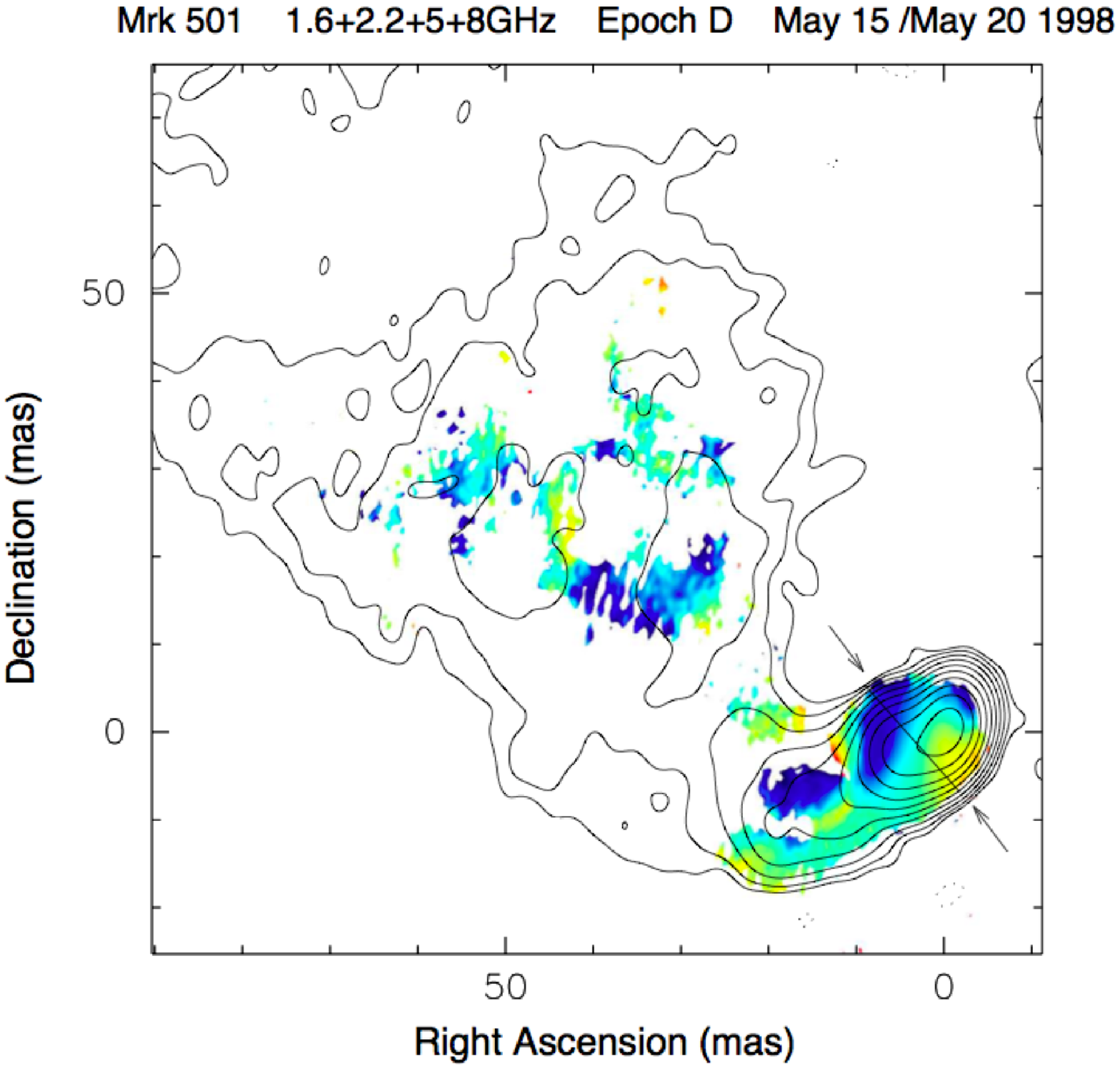}}
\caption{Example of a transverse Faraday-rotation gradient detected
  across the jet of the blazar Mrk 501; the contours show the distribution of the radio intensity at 18cm, and
the colour scale the distribution of the observed Faraday rotation
\citep{CrokeOSullivanGabuzda2010}. The redshift of this object is
0.034, so that 1 mas is 0.66 pc.}
\label{fig-mrk501-rm}
\end{figure}

Nevertheless, firm detections of transverse Faraday rotation gradients have
now been made for a number of AGNs.
The first report of the detection of a transverse Faraday gradient across an
AGN jet, interpreted as evidence for a helical jet $B$ field, was made by
\cite{Asada2002}, based on Very Long Baseline Array (VLBA) observations of the
nearby AGN 3C\,273; this was later confirmed by \cite{ZavalaTaylor2005}, and
the evolution of the gradient analyzed by \cite{Asada2008a}. In the
meantime, transverse Faraday-rotation gradients in several more objects were 
presented by \cite{GabuzdaMurrayCronin2004}. There have since then been
at least ten further papers in refereed journals reporting the 
presence of transverse Faraday-rotation gradients across the jets of more than
a dozen AGNs on parsec
scales, interpreted as evidence that these jets carry helical magnetic fields
\citep{Asada2008a, Asada2008b, Asada2010, Gabuzda2008, Gomez2008,
OSullivanGabuzda2009RM, Kharb2009, MahmudGabuzdaBezrukovs2009,
CrokeOSullivanGabuzda2010, Kronberg2011}; one example is shown in 
Fig.~\ref{fig-mrk501-rm}. A number of other tentative cases
have been presented in conference proceedings or identified in
previously published Faraday-rotation images in the literature (e.g., 
\citealt{Contopoulos2009,ZavalaTaylor2003,ZavalaTaylor2004}). 

This suggests that, when the observing conditions enable their detection, 
transverse Faraday-rotation gradients across AGN jets are not uncommon on
parsec scales. This result is potentially of crucial importance for our 
understanding of AGN jets, since it provides direct evidence that these jets
carry helical magnetic fields, which, in turn, has obvious implications for
the jet-launching mechanism. In addition, the presence of an appreciable
ordered toroidal field component would imply that the jets carry current. 
It is therefore important to try to ensure that the requirements for the
reliable detection of transverse Faraday-rotation gradients are well understood
and applied.
 
It has sometimes been suggested that apparently transverse Faraday-rotation
gradients could come about due to gradients in the ambient electron density,
and therefore not have anything to do with the presence of a helical jet $B$ field.
An interesting point here is that, if the viewing angle of the jet in the jet
rest frame is not too far from $90^{\circ}$ (the jet is viewed roughly from the
side in the jet rest frame), the observed Faraday rotation associated with the
jet's helical field should change sign across the jet; if the jet viewing angle
in the jet rest frame is not close to  $90^{\circ}$, the jet's helical field
will still give rise to a transverse Faraday-rotation gradient, but there may
not be a change in the sign of the Faraday rotation across the jet. Therefore,
the clear detection of a change in the sign of the observed Faraday rotation
across an AGN jet can easily be explained by the presence of a helical field,
but not by electron-density gradients; such sign changes have been observed
for a number of AGN jets (e.g., \citealt{Asada2008b, GabuzdaMurrayCronin2004, 
Gabuzda2008, MahmudGabuzdaBezrukovs2009}).

Other observational problems that could potentially hinder the detection of
transverse Faraday-rotation gradients associated with helical magnetic fields,
or distort their interpretation, include the limited resolution available to
centimeter-wavelength VLBI observations and optical depth effects in the region
of the observed VLBI core \citep{TaylorZavala2010}. The presence of non-monotonic 
and/or complex transverse Faraday-rotation gradients in the core region have 
been directly demonstrated by the simulations of \cite{BroderickMcKinney2010} 
and \cite{Porth/Fendt:2010}. These simulations indicate the difficulty of accurately
deriving parameters of the jet and the helical field it is believed to carry 
based on observed transverse Faraday-rotation gradients. Note, however, that 
the direction of the simulated gradients is correct when they are monotonic; 
since most observational studies carried out thus far have focused on the detection 
of {\em monotonic} transverse gradients and their direction, the results of these 
studies are not greatly affected by the difficulties indicated by the simulation 
results.

\cite{TaylorZavala2010} have proposed that a transverse Faraday-rotation gradient
should span three ``resolution elements'' (taken to correspond to three beamwidths)
in order for the gradient to be considered reliable; however, no theoretical basis
was provided for this criterion.  It is therefore of interest to develop more 
systematic and objective approaches 
to determining the resolution necessary to reliably distinguish transverse 
Faraday-rotation gradients, and results from a number of such studies have 
begun to appear. For example, \cite{MurphyGabuzda2011} have calculated
transverse intensity, polarization and Faraday-rotation distributions and
subjected them to convolution with beams of various sizes. The initial results 
of these studies
indicate that the detection of transverse polarization structure associated 
with helical jet $B$ fields is much more robust to convolution with beams comparable
to the jet width than is transverse intensity structure, although, of course,
the convolution somewhat distorts the intrinsic polarization structure.
Similarly, convolution of transverse Faraday-rotation gradients with beams 
that are comparable to or even larger than the transverse size of the jet 
reduces the magnitude of the Faraday-rotation gradient, but does not destroy the 
gradient completely.  \cite{Hovatta2012} have recently presented the results of
four-frequency VLBI polarization observations of roughly 200 AGNs, together with
Monte Carlo simulations modeling the polarization uncertainties and how they
are manifest in Faraday-rotation maps. These Monte Carlo simulations suggest
that transverse Faraday-rotation gradients spanning as little as 1.5 beamwidths 
can be reliable, provided that the difference between the Faraday rotations on
either side of the jet exceeds $3\sigma$. 

These results are also consistent with the general relativistic
MHD simulations of \cite{BroderickMcKinney2010}, which directly demonstrate the
presence of transverse Faraday-rotation gradients across regions that are only 
marginally resolved: transverse structures in their simulated Faraday-rotation 
distributions with intrinsic sizes less than 0.05~mas remain clearly visible even 
when convolved with a 0.9-mas beam. All this suggests that the minimum resolution 
required to reliably detect a transverse Faraday-rotation gradient across an AGN 
jet is considerably less than the three-beamwidth criterion suggested by 
\cite{TaylorZavala2010}, provided that the Faraday rotation measures 
measured on opposite sides of the jet can reliably be demonstrated to  
differ at more than the $3\sigma$ level. Of course, this concerns only
the detection of the presence and direction of a transverse Faraday-rotation 
gradient; reliable measurement of the intrinsic magnitude of such a gradient 
would require much higher resolution. 

Another issue that is currently being investigated is the question of how to
accurately estimate the uncertainties at specific locations in VLBI images; this 
is non-trivial due to the complex computational procedures undertaken in the 
mapping process and the strong correlation between neighbouring pixels, due 
to convolution with the CLEAN beam.  In the past, it has been usual to assign 
the uncertainty for a pixel in images of the Stokes parameters $I$, $Q$ and 
$U$ to be equal to the root-mean-square deviation about the mean calculated for 
a large region far from areas of source emission, but this is certainly an 
underestimate. \cite{Hovatta2012} and \cite{Mahmud2012} use Monte Carlo 
simulations to address this issue, motivated by the need to derive accurate 
estimates of uncertainties for individual locations in Faraday-rotation images. 

Another interesting question in connection with the detection of transverse
Faraday-rotation gradients across AGN jets is whether it is possible to
observationally distinguish between a helical and a toroidal magnetic field.
The point here is that the detection of the transverse gradient only demonstrates
the presence of an ordered toroidal field component, without providing any
information about the presence of a longitudinal field component.  
Both helical and toroidal jet $B$ fields could give rise to ``spine+sheath'' 
transverse polarization structures, and a rise in the degree of polarization 
toward the edges of the jet, as is commonly observed. One means of distinguishing 
between these two 
possibilities is provided by the transverse intensity and polarization 
structure of the jet: a purely toroidal field can only 
produce symmetrical transverse intensity and polarization structures (e.g.
\citealt{Zakamska2008}), whereas
a helical field can produce either symmetrical or asymmetrical transverse
structures, depending on the viewing angle of the jet and the pitch angle of
the helical field (e.g., \citealt{MurphyGabuzdaCawthorne2010, MurphyGabuzdaCawthorne2012}). Therefore, the joint
detection of a transverse Faraday-rotation gradient and clearly asymmetrical
transverse polarization structure would provide clear evidence for a helical,
rather than a purely toroidal, jet $B$ field.  
 
Finally, note that Faraday rotation associated with detected transverse
gradients observed can be external (not occurring throughout the radiating 
volume of the source), if it is associated with regions of helical magnetic 
field in the outer layers of the jet, or in the immediate 
vicinity of the jet.
 
\subsection{Evidence for reversals in the direction of the toroidal field
component}
\label{dcg:reversals}

One of the most unexpected results to come out of recent parsec-scale
Faraday-rotation studies is the detection of reversals in the direction
of transverse Faraday-rotation gradients across AGN jets, both with distance
from the base of the jet \cite{MahmudGabuzda2008, HallahanGabuzda2008, 
Mahmud2012} and with time 
\citep{MahmudGabuzdaBezrukovs2009}. This is at
first glance difficult to understand, since the direction of the observed
Faraday-rotation gradient is determined by the direction of the toroidal
field component, which is, in turn, essentially determined by the direction
of rotation of the central black hole and accretion disc, together with the
direction of the net poloidal component of the initial ``seed'' field that
is wound up. It is not physically plausible for the direction of the
system's rotation to change on measureable time scales, and changing the 
direction of the jet poloidal component of the seed field would seem to 
require a change in the polarity of the field of the black hole, which 
likewise, seems implausible, at last on measureable time scales. 
\cite{BisnovatyiKogan2007} has suggested the possiblity that, under
certain circumstances, torsional oscillations can develop in a jet, which could
give rise to changes in the direction of the toroidal $B$-field component;
in this scenario, the direction of the observed transverse Faraday-rotation 
gradients would reverse from time to time when the direction of the 
torsional oscillation reverses, and this reversal pattern would presumably 
then propagate outward with the jet. 

However, \cite{MahmudGabuzdaBezrukovs2009} and \cite{Mahmud2012} have 
suggested that a simpler and more likely explanation
is a magnetic-tower-type model, with poloidal magnetic flux and poloidal
current concentrated around the central axis \citep{Lynden-Bell:1996,
Nakamura2006}. Fundamental physics dictates that the magnetic-field
lines must close; in this picture, the magnetic field forms meridional
loops that are anchored in the inner and outer parts of the accretion disc,
which become twisted due to the differential rotation of the disc.
This should essentially give rise to an ``inner'' helical $B$ field near the
jet axis and an ``outer'' helical field somewhat further from the jet axis.
These two regions of helical field will give rise to oppositely directed
Faraday-rotation gradients, and the total observed gradient will be determined 
by which region of helical field dominates the observed Faraday rotation. 
The presence of a change in the direction of the observed 
transverse Faraday-rotation gradient with distance from the jet base
could represent
a transition from dominance of the inner to dominance of the outer helical
$B$ fields in the total observed Faraday rotation. The reversals of the
Faraday-rotation gradients with time observed for 1803+784 
\citep{MahmudGabuzdaBezrukovs2009} 
could come about if the physical conditions in the jet changed such
that there was a change in whether the inner or outer region of helical
$B$ field dominated the total observed Faraday rotation. 
If this interpretation is correct, these observations of reversals of the 
direction of transverse Faraday-rotation gradients across AGN jets 
represent the first observational evidence that the jet magnetic field
closes in the outer accretion disc. 

\subsection{Evidence for the action of a cosmic battery}
\label{dcg:PRbattery}

Any transverse Faraday-rotation gradient can be described as being directed
either clockwise (CW) or counter-clockwise (CCW) on the sky, relative to the
base of the jet.  \cite{Contopoulos2009} have reported a significant 
excess of CW transverse Faraday-rotation gradients for parsec-scale AGN
jets, based on transverse Faraday-rotation gradients identified in maps
from the literature. Considering the gradient closest to the VLBI core if 
two distinct regions with transverse gradients were present (as is
described in the previous subsection) yielded 29 transverse Faraday-rotation 
gradients on parsec scales, of which 22 were CW and only 7 CCW, with the 
probability of this coming about by chance being less than 1\%.
This is an extremely
counterintuitive result, since the direction of the helical field threading
an AGN jet should essentially be determined by the direction of rotation 
of the central black hole and accretion disc, together with the direction 
of the poloidal ``seed'' field that is wound up, and our instincts tell us 
that both of these should be random. 

\begin{figure}
\epsfxsize\textwidth
\epsfbox{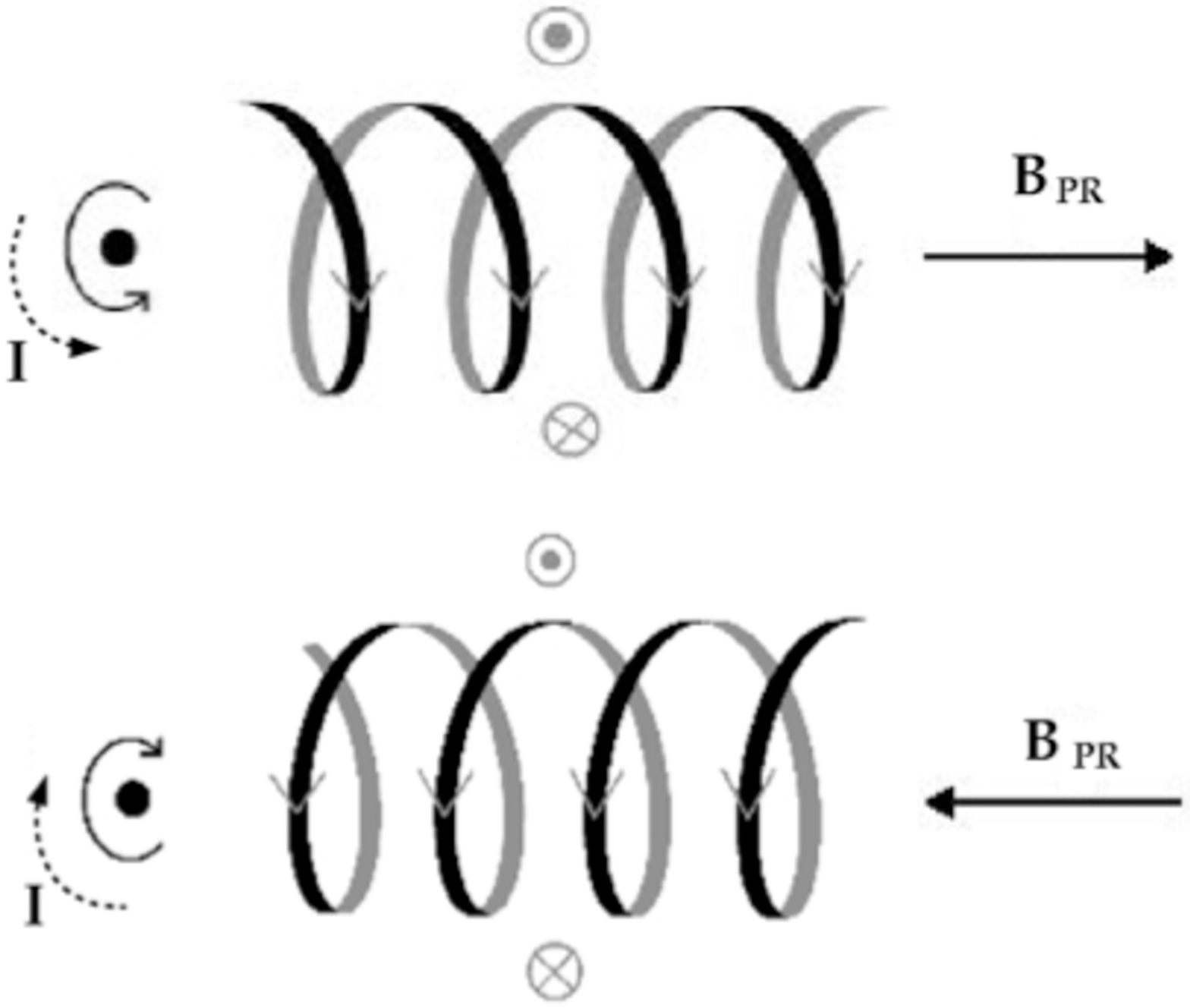}
\caption{Illustration of how the action of the Poynting--Robertson battery
couples the direction of rotation of the accretion disc and the direction
of the poloidal seed field that is wound up, leading to the creation of
Faraday-rotation gradients associated with the resulting helical $B$ field that
are CW on the sky, independent of the direction of disc rotation 
\citep{Mahmudthesis}.}
\label{fig-PR-schematic}
\end{figure}

\cite{Contopoulos2009} suggest that
this seemingly bizarre result can be explained in a straightforward way via
the action of a mechanism they call the ``Poynting--Robertson cosmic battery.''
The essence of this mechanism is the Poynting--Robertson drag experienced by
charges in the accretion disc, which absorb energy emitted by the active
nucleus and re-radiate this energy isotropically in their own rest frames. 
Because these charges are rotating with the accretion disc, this radiation
will be beamed in the forward direction of their motion, i.e., in the direction
of the disc rotation. Due to conservation of momentum, the charges then
feel a reaction force opposite to the direction of their motion; since the
magnitude of this force exhibits an inverse dependence on the mass of the
charge, this leads to a difference in the deceleration experienced by the
protons and electrons in the disc. Since the electrons are decelerated
more strongly, this leads to a net current in the disc, in the direction of
rotation. This current, in turn, gives rise to a net poloidal magnetic field
whose direction is coupled to the direction of the current in the disc, i.e.,
to the direction of the disc rotation. This coupling of
the disc rotation and the direction of the poloidal field that is ``wound 
up'' breaks the symmetry in the direction of the toroidal field component,
and predicts that the observed Faraday-rotation gradients should be
predominantly CW on the sky, independent of the direction of the disc rotation
as seen by the observer (Fig.~\ref{fig-PR-schematic}). In a 
``nested helical field'' type picture such
as that described in the previous subsection, the inner/outer regions of helical
field should give rise to CW/CCW Faraday-rotation gradients; therefore, the
observed excess of CW implies that the inner region of helical $B$ field
dominates on parsec scales.
 
If this excess of CW Faraday-rotation gradients is confirmed by further
studies, this will have cardinal implications for our understanding of
AGN jets. The development of approaches to carrying out objective searches
for reliable transverse Faraday-rotation gradients for as many AGNs as 
possible is of interest in this connection.
Independent of the physical origin of the predominance of a 
particular orientation, the results of \cite{Contopoulos2009}
essentially rule out the possibility
that most of the observed gradients are spurious, and not associated with
the toroidal field components of the jets, since spurious (incorrectly
identified) gradients should certainly be randomly oriented on the sky. 
If the Poynting--Robertson battery does not operate sufficiently efficiently
to provide the observed excess of CW gradients, then some other mechanism
that can couple the direction of rotation of the accretion disc and the
direction of the poloidal field that is wound up by differential rotation
must be identified.

\subsection{Connection with kiloparsec scales}
\label{dcg:kpc-jets}

Recently, \cite{Christodoulou2012} (see also \citealt{Gabuzda2012}) searched 
the literature for
transverse Faraday-rotation gradients across the kpc-scale jets
and lobes of AGNs and radio galaxies. This yielded six AGNs displaying
continuous, monotonic RM gradients across their jets, oriented roughly 
orthogonal to the local jet direction.  The most fundamental implication of 
this result is that transverse Faraday-rotation gradients that can most
naturally be interpreted as evidence for toroidal or helical magnetic
fields threading the jets are also present on kpc scales. This is 
an important result, because, if
the jets of AGNs carry helical magnetic fields, these should be present on
essentially all scales where the jets propagate, provided that the intrinsic
field structure of the jet is not disrupted by interactions with the 
surrounding environment.

Another intriguing aspect of the analysis of \cite{Christodoulou2012} is
that all six of the firm transverse kpc-scale Faraday-rotation gradients
identified are oriented CCW on the sky, relative to their jet bases. 
This is on the verge of being a firm result statistically: based on a simple
unweighted binomial probability function, the probability for six out of six
gradients to be CCW by chance is about 1.5\%. However, the significance of
this result is tentatively supported by VLBA observations of 3C\,120 
\citep{Coughlan2010}, Mrk501 \citep{CrokeOSullivanGabuzda2010}, and 
1749+701 \citep{HallahanGabuzda2008}: in each case, transverse CCW 
Faraday-rotation gradients are found on scales of tens of pc. Further, 
all AGNs for which reversals of their transverse Faraday-rotation gradients 
have been observed display CCW Faraday-rotation gradients further from the jet 
base \citep{MahmudGabuzda2008, Mahmud2012, ReichsteinGabuzda2010, Coughlan2010}.
This growing evidence for a predominance of CW/CCW Faraday-rotation gradients
on parsec/kpc scales
 can be interpreted as a consequence of a nested-helical-field
configuration, as is described above, with the inner region of helical
field dominating on parsec scales and the outer region of helical field
dominating on kpc scales.  
This picture is consistent with theoretical studies of the jet launching
mechanism \citep{BlandfordPayne1982, ContopoulosLovelace1994, Contopoulos1995,
Spruit2010}, which suggest that the field
is effectively wound up only beyond the Alfv\'en distance,
which is $\sim$10 times the radial extent of the outflow at its base.
Thus, the theoretical models suggest that the magnetic field in the
inner jet and the outer extended accretion disc will develop significant 
toroidal components on distances beyond $\sim$10~AU and $\sim$10~pc,
respectively, from the AGN centres; this would predict a predominance
of CW Faraday-rotation gradients on observed scales less than  $\sim$10~pc and
of CCW gradients on scales greater than  $\sim$10~pc, consistent with the
results of \cite{Contopoulos2009} and \cite{Christodoulou2012}. Thus, although
the results of \cite{Christodoulou2012} must clearly be confirmed based on
a larger number of AGNs, they have the potential to considerably influence
our understanding of the overall magnetic-field configurations of AGN jets.

The analysis of \cite{Christodoulou2012} suggests that the 
occurrence of reliably detectable transverse Faraday-rotation gradients
is appreciably lower on kpc than on parsec scales: they were 
able to find reasonably clear transverse Faraday-rotation gradients in only 
six of about 85 objects.  In fact, this is quite natural.  There is an
appreciable turbulent, inhomogeneous component to the thermal ambient
media surrounding the jets on kpc scales, which superposes a more or
less random pattern over the systematic pattern due to the helical fields
(see the discussion in Section~\ref{mjh:kpc-jet}).
This random component in the Faraday-rotation distribution may dominate in 
the majority of cases. Thus, it is natural that most of the 
observed Faraday-rotation distributions
appear random and patchy, but the overall pattern due to the helical fields
sometimes comes through. This suggests that, on average, it is easier
to detect the systematic Faraday-rotation
component due to helical jet magnetic fields on
parsec scales, where the ordered inner field is more dominant. 

The additional complications imposed by the large-scale interaction
with the jet and the ambient medium, and the structures (lobes,
hotspots etc.) generated by this interaction, are discussed in the
following Section.

\section{Interaction with surrounding medium and termination}

\subsection{Introduction}
\label{mjh:intro}
The best-studied regions of jet termination and the best examples of
interactions of jets with the surrounding medium are furnished by
radio-loud AGN, The parsec-scale AGN jets discussed in Section 3 generally
feed into kpc-scale structures --- kpc-scale jets, plumes, lobes and
hotspots. The magnetic fields in these structures must exhibit some
continuity with the fields set by the jet launching process (Section
2) and exhibited on the parsec scale (Section 3), particularly in the
inner parts of kpc-scale jets. However, on the hundred-kpc scales of
the jet termination and the downstream structures such as lobes and
plumes, it seems likely that much of the initial structure has been
erased. As is the case on parsec scales, the intrinsic polarization of
synchrotron emission gives us information about the direction and
degree of ordering of the magnetic field, though we always need to
bear in mind that what we can measure is a projected,
polarized-emission-weighted line-of-sight average of the intrinsic
degree of polarization and the $E$- or $B$-vector angle. In addition,
Faraday rotation both external and internal to the emitting medium has
an effect on what we observe. A good deal is known about the
polarization properties of radio-loud AGN on the largest scales, and
we summarize the observational situation in Sections
\ref{mjh:kpc-jet}, \ref{mjh:hotspot} and \ref{mjh:lobe}. We then go on
to discuss the relationship between observations and modelling in
Section \ref{mjh:modelling} and the situation in the comparable
structures in stellar-scale outflows in Section \ref{mjh:stellar}.
Before that, however, we discuss the methods by which magnetic field
strength can be measured in AGN on these scales.

\subsection{Measuring magnetic field strengths on the kpc scale}
\label{mjh:strength}

Unfortunately, synchrotron radiation on large scales can tell us
little or nothing about the magnetic field {\it strength}, although
this is clearly a key piece of information in understanding the
dynamics and energetics of the jet and its environment, without
significant (and possibly incorrect) additional assumptions. The
methods discussed in Section \ref{dcg:strength} can only be applied in
regions with a well-known structure and a non-negligible optical
depth, and kpc-scale structures are almost always optically thin at
accessible frequencies.

The standard equipartition/minimum energy assumptions
\citep{1956ApJ...124..416B} are widely used, but even at best only provide us with a
plausible order of magnitude estimate of the magnetic field strength: there is
no a priori reason to suppose that the plasma reaches the minimum
energy or equipartition conditions, and even if it does, these assumptions
require us to guess the appropriate values for the filling factor and
the fraction of energy in non-radiating particles, $\phi$ and $\kappa$
in the standard notation for equipartition:
\begin{equation}
{{B^2}\over{2\mu_0}} = {{1+\kappa}\over{\phi}} \int E N(E){\rm d}E
\label{eq:equip}
\end{equation}
Finally, long-standing practice in the radio-galaxy community is to
carry out the integral in eq.\ (\ref{eq:equip}) over a fixed range in
{\it observed frequency}, not in electron energy, although the latter
seems more physically reasonable \citep{1985ApJ...291...52M,1997A&A...325..898B,1998MNRAS.294..615H} and the differences between these two sets of assumptions
can lead to significantly different results \citep{2005AN....326..414B}.

If equipartition and minimum energy are unsatisfactory, what else is
available? Another widely used approach that gives at least some
constraints is to consider the properties of a medium external to the
radio source, which may be easier to measure. For example, thermal
bremsstrahlung in the X-ray from the environments of large-scale jets
and lobes can give us measurements of the temperature and density, and
therefore pressure, of the external environment. On the reasonable
assumption that the pressure in a radio-emitting component cannot be
much less than that in the medium in which it is embedded, we
therefore have a constraint on the total energy density in the source.
Combining this with the synchrotron measurements, we can derive
values for the magnetic field strength. However, these are not unique
(for a given pressure greater than the minimum pressure there are two
possible values of $B$ to choose from) and the method is still dependent on
assumptions about $\kappa$ and $\phi$ as well as field geometry.

By far the best method of measuring magnetic field strength in
extended components of radio-loud AGN is the use of inverse-Compton
emission. The volume emissivity from the inverse-Compton process
depends essentially on the electron and photon number densities as a
function of energy. If the properties of the photon field are known, a
detection of inverse-Compton emission gives us a constraint on the
normalization of the electron energy spectrum. We can use this in
combination with a measurement of synchrotron emissivity to estimate
the $B$-field in a very robust way: assuming we know the geometry from
radio observations, the only model dependence comes from our
assumptions about the shape of the electron energy spectrum, since
typically we will be observing energetically different electron
populations using the two processes. For very simple spatial and
spectral properties of the electron populations, analytical estimates
of $B$ from inverse-Compton detections can be derived
\citep[e.g.][]{1979MNRAS.188...25H}; for more complex situations,
numerical codes must be used to integrate the relevant equations
\citep[e.g.][]{1998MNRAS.294..615H,2001A&A...372..755B,2002ApJ...581..948H}.

Inverse-Compton emission from large-scale components of radio-loud AGN
was first discovered in the X-ray
\citep{1994Natur.367..713H,1995ApJ...449L.149F} and X-ray observations
dominate the current applications of this technique of measuring $B$.
A complication of this is that various other processes, including
thermal bremsstrahlung from the hot phase of the IGM and synchrotron
emission from high-energy electrons, can produce X-ray emission from
the radio sources and their environments. In the following subsections we
will discuss the available constraints on magnetic field strengths
using this method, bearing in mind these observational limitations.

\subsection{Magnetic fields in kpc-scale jets}
\label{mjh:kpc-jet}

Kpc-scale jets are often strongly polarized, as discussed in Section
\ref{dcg:kpc-jets}. In the most powerful jets, the apparent magnetic
field direction after correction for Faraday rotation is normally
along the jet, often following bends in the jet very closely
\citep[e.g.][]{1994AJ....108..766B}. In the less powerful jets of FRI
radio galaxies, there is often a transition between a parallel field
configuration in the inner few kpc to a perpendicular field in the
centre of the jet further out: parallel fields may be observed at the
edge of the jet \citep[e.g.][]{1984ARA&A..22..319B}. These FRI jets
are the jets that have been studied in the greatest detail to date,
because their proximity and relative brightness allows sensitive
measurements to be made in regions of the jet that are transversely
resolved on scales from sub-arcsec to tens of arcsec (hundreds of pc
to a few kpc, at typical distances). In addition, both a jet and
counterjet are routinely detected in FRIs, and the assumption that the
jet and counterjet are intrinsically symmetrical is a viable one,
allowing the effects of relativistic beaming and aberration to be
separated from the intrinsic (rest-frame) behaviour of the source.
Detailed modelling of these FRI jets
\citep[e.g.][]{2002MNRAS.336..328L,2006MNRAS.372..510L} shows that the
observed polarization structures can be modelled in terms of a
magnetic field that is locally random but anisotropic. The anisotropic
component of the field is mainly a combination of toroidal and
longitudinal components, with the toroidal component dominating both
at the edges of the jets and at large distances from the nucleus.
Globally ordered helical models for the magnetic field (cf.\ Section
\ref{dcg:helical}) have been ruled out in some of the best-studied
cases \citep{2005MNRAS.363.1223C,2006MNRAS.372..510L}: if the field is
initially helical, it must evolve away from such a configuration by
the kpc scale.

Measurement of $B$-field strengths in jets is more difficult; although
resolved X-ray emission from jets is common (see e.g. \citealt{2006ARA&A..44..463H} for a review), there are two fairly serious problems
in using them to determine $B$. The first is that some of the X-ray
emission seen from these structures, particularly in low-power
objects, is unequivocally known to be synchrotron in origin \citep[e.g.][]{2001MNRAS.326.1499H}. Where synchrotron emission is dominant, we
only have a (usually uninteresting) lower limit on $B$ in the jet. The
second is that the jets of the most powerful objects are known to be
at least mildly relativistic on kpc scales \citep{1994AJ....108..766B,1997MNRAS.286..425W,1999MNRAS.304..135H,2009MNRAS.398.1989M} and
this introduces additional model dependence into any estimates of $B$,
since the observed properties of the jet now depend on the bulk
Lorentz factor $\Gamma$ and the angle to the line of sight $\theta$. A
good deal of interest was triggered by the discovery that the X-ray
emission from the powerful source PKS 0637$-$752 \citep{2000ApJ...540L..69S}, which could not be explained by a one-zone synchrotron
model, could be explained as inverse-Compton scattering of the CMB
with a magnetic field close to equipartition, using a Lorentz factor
($\Gamma \sim 10$) and angle to the line of sight close to those
implied by VLBI observations in the nucleus
\citep{2000ApJ...544L..23T,2001MNRAS.321L...1C}, which would imply no jet deceleration from pc to
hundred-kpc scales. However, this certainly does not seem to be the
case for all such powerful quasars \citep{2006MNRAS.366.1465H} and later work on
the broad-band SED of the knots in the best-studied example of this
class, 3C\,273, suggests that the data are inconsistent with an
inverse-Compton model anyway \citep{2002A&A...385L..27J,2005A&A...431..477J}. As yet there
is therefore no likelihood of being able to determine large-scale
magnetic fields from inverse-Compton emission in the jets of the most
powerful objects, though in (bright, nearby) lower-power objects we
may hope for inverse-Compton emission at higher energies (see Section
\ref{mjh:strength-progress}).

\begin{figure}
\epsfxsize\textwidth
\epsfbox{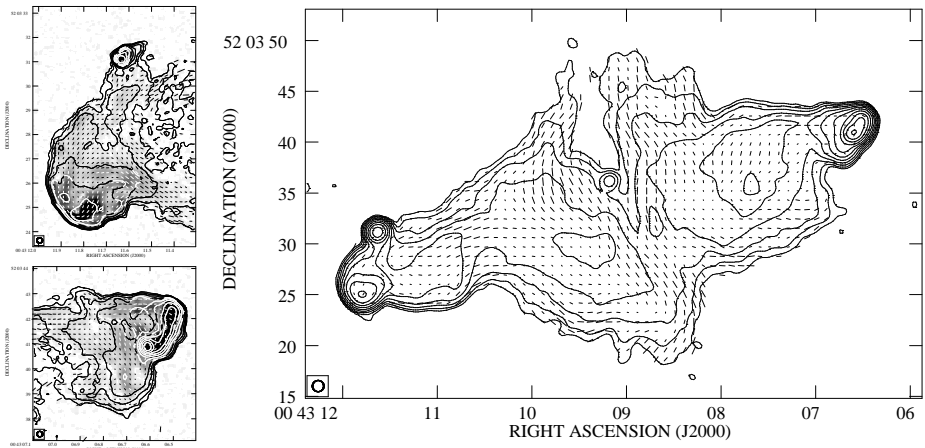}
\caption{Complex polarization structure in the lobes and hotspots of
  the FRII radio galaxy 3C\,20 at 8 GHz. Right: the overall structure
  of the source. Top left: the E hotspot. Bottom left: the W hotspot.
  Polarization vectors have length representing fractional
  polarization and are plotted at $90^\circ$ to the direction of the
  polarization $E$-vector, thus giving a representation of the
  magnetic field direction. Contours show total intensity, increasing
  logarithmically by a factor of 2 at each step. The greyscales shown
  for the hotspots indicate polarized intensity to give an idea of the
  complex filamentary structure that is visible in polarized emission.
  Data from \cite{1997MNRAS.288..859H}.}
\label{3C20-fig}
\end{figure}

\subsection{Jet termination}
\label{mjh:hotspot}

\begin{figure}
\epsfxsize\textwidth
\epsfbox{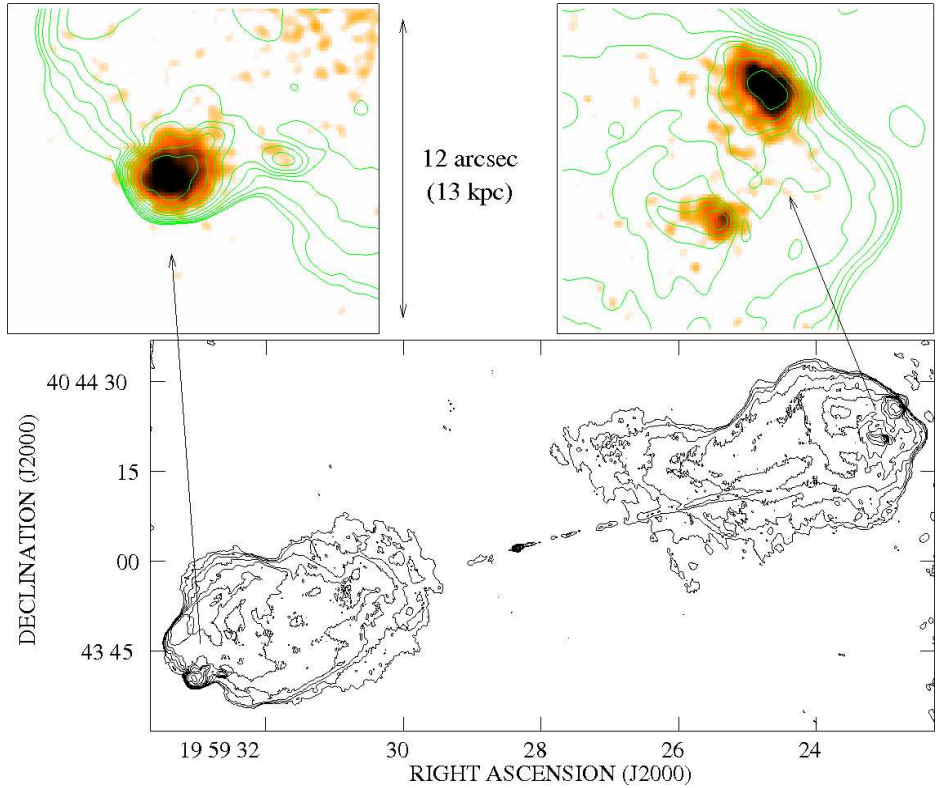}
\caption{Synchrotron self-Compton X-ray emission from the hotspots of
  Cygnus A (3C\,405), as first reported by \cite{1994Natur.367..713H}. Lower panel: contours show the 4.5-GHz radio emission from
  the whole source, from the image of \cite{1991ApJ...383..554C}. Upper
  panel shows the same contours overlaid on smoothed {\it Chandra}
  data in the 0.5-5.0 keV range from the dataset of \cite{2006ApJ...644L...9W}. Note the close morphological agreement between radio
  and X-ray emission. The level of the inverse-Compton emission
  implies field strengths close to the equipartition value in the hotspots.}
\label{3C405-fig}
\end{figure}

The bright compact regions at the terminations of powerful jets, known
as hotspots, are believed to be shocks at which the jet material
interacts with plasma already present in the lobes. They are well
studied in the radio because of their high surface brightness, and
polarization observations at high resolution
\citep[e.g.][]{1999AJ....118.2581C} show them to have complex
polarization structures (e.g.\ Fig.\ \ref{3C20-fig}), presumably
because of the complex hydrodynamics in this part of the source. There
is a general tendency for the inferred magnetic field angle to be
perpendicular to the jet direction on entry to the hotspot
\citep[e.g.][]{1997MNRAS.291...20L,1997MNRAS.288..859H}, possibly
because of field compression at the shock.

Because the hotspots are often the brightest compact features in
sources in which they occur, they would be expected {\it a priori} to
be good sources of so-called synchrotron self-Compton (SSC) emission,
where synchrotron emission provides the seed photons for
inverse-Compton scattering (Fig.\ \ref{3C405-fig}). Observational
constraints suggest that the bulk flow through hotspots is at best
mildly relativistic \citep[e.g.][]{2008MNRAS.390..595M} and so beaming
is less important. Early observations of X-ray emission from hotspots
were consistent in many cases with the X-rays being SSC and the
magnetic field strengths implied being close to equipartition
\citep{1994Natur.367..713H,2000ApJ...530L..81H,2001MNRAS.323L..17H,2002ApJ...581..948H}.
However, at the same time, a number of hotspots showed X-ray emission
that was much brighter than the expectation from SSC at equipartition
and (in some cases) was spectrally inconsistent with being SSC at all;
this X-ray emission seemed likely to be partly or wholly synchrotron
in origin \citep[e.g.][]{1998ApJ...499L.149H,2001ApJ...547..740W}.
\cite{2004ApJ...612..729H}, based on the large amount of hotspot data
collected in the first few years of the {\it Chandra} mission, showed
that it was plausible that the difference was controlled by the
luminosity, and hence the magnetic field strength, of the hotspots
\citep[cf.][]{2003MNRAS.345L..40B}; the most luminous hotspots have
magnetic field strengths high enough that strong synchrotron losses
inhibit the acceleration of electrons to energies at which X-ray
synchrotron emission would be detected. If this is the case, then it
is plausible that all hotspots, not just the subset for which SSC has
been detected, have magnetic fields close to the equipartition value
\citep{2004ApJ...612..729H,2005ApJ...622..797K}. Remarkably,
therefore, it seems that there is a rapidly-acting (since hotspots are
short-lived), small-scale physical process that acts to amplify
magnetic field strengths up to values within a factor of a few of their
equipartition values. These field strengths may be up to 20 nT (200
$\mu$G) in the brighest sources.

\subsection{Lobes}
\label{mjh:lobe}

Beyond the end of the jet and the jet termination shock, if present,
are the structures fed by the jet. The most common structure seen is
large-scale lobes of synchrotron-emitting material extending back
towards the centre of the host galaxy, although extended diffuse
plumes are seen in a minority of low-power radio galaxies.

The lobes are often strongly polarized, with the typical apparent
magnetic field direction being parallel to contours of constant total
intensity surface brightness (e.g.\ Fig.\ \ref{3C20-fig}). Because of
the large spatial scales, strong polarization and large sizes of
magnetic field structures, lobes are the best places for detailed
studies of Faraday rotation intrinsic to the source or its local
environment. The change in polarization position angle $\Delta\theta$
produced by a foreground Faraday screen is given by $\Delta\theta =
RM\,\lambda^2$, where $RM$ is the rotation measure,
\begin{equation}
RM = C\int n_e\, {\mathbf B}\cdot {\rm d}{\mathbf l}
\end{equation}
Thus in principle, with multi-wavelength observations to give the
required $\lambda^2$ coverage, we can measure both $RM$ and the
intrinsic polarization angle and so constrain the electron density and
field strength in the Faraday screen. In practice, there are a number
of complications \citep[e.g.][]{1984pete.conf...90L}. Faraday-active material (thermal
electrons) {\it inside} the synchrotron-emitting region will induce
depolarization and departures from a $\lambda^2$ law; in fact, the
absence of this effect gives very strong, albeit model-dependent,
limits on the possible densities of thermal electrons inside the lobes
\citep[e.g.][]{1987ApJ...316..611D}. Even if the Faraday screen is external to
the lobes, its characteristic size scale must be fully resolved by the
radio observations, otherwise beam depolarization results and the
problem of determining $n_e$ and $B$ becomes much harder (though in
some circumstances it is possible, see e.g.\ \citealt{2010MNRAS.401.2697H}).
When Faraday rotation external to the radio galaxy is measured, the
derived $RM$ values are consistent with being due to the hot phase of
the intracluster medium, and with constraints on the external density
the field strengths may be estimated. Observed $RM$ values range from
$\pm 100$ rad m$^{-2}$ in typical radio galaxies to $\pm 4000$ rad
m$^{-2}$ in Cygnus A, at the centre of a rich cluster \citep{1987ApJ...316..611D} and the estimated energy densities in magnetic fields are
then comparable to the thermal energy densities of the plasma. The
structures observed in Faraday rotation may also be used to constrain
the power spectrum of magnetic field fluctuations in the external medium \citep{2008MNRAS.391..521L}.
However, in the absence of {\it internal} Faraday rotation, for which
there is no conclusive evidence in any object on kpc scales, these
estimates of $B$ tell us nothing about the field strength or its
variations in the lobes themselves.

\begin{figure}
\epsfxsize\textwidth
\epsfbox{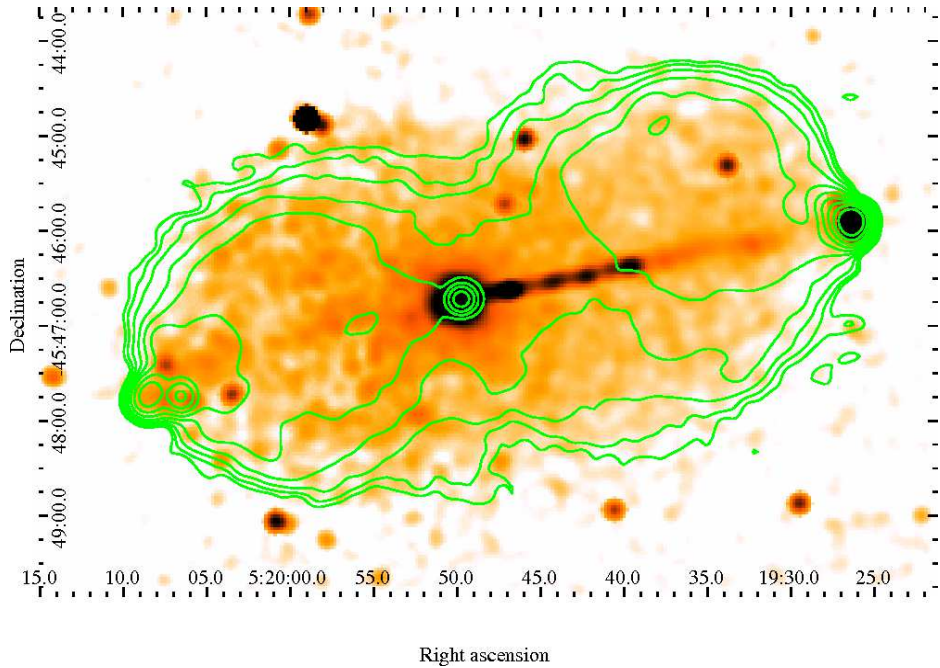}
\caption{X-ray emission from the nearby FRII radio galaxy Pictor A.
  The colour scale shows Gaussian-smoothed {\it Chandra} counts in the 0.5-5.0 keV
  range, using the data described by \cite{2010ApJ...714L.213M}: contours are from the 1.4-GHz radio map of \cite{1997A&A...328...12P}. Note the X-ray emission (most likely synchrotron, see
  \citealt{2005MNRAS.363..649H}) from the jet and hotspot, the excellent
  match between the boundaries of the X-ray and radio emission, and the
  relatively constant X-ray surface brightness of the lobes, in
  contrast to the strongly varying radio surface brightness (contours
  are logarithmic, increasing by a factor 2 at each step).
}
\label{fig-pica}
\end{figure}

The lobes are expected to be particularly good sources of inverse-Compton
emission from scattering of diffuse photon populations such as the CMB
and the extragalactic background light (Fig.\ \ref{fig-pica}). Because their geometry and
synchrotron properties are also relatively well constrained, because
it is certain that they do not have relativistic bulk motions, and
because there is no evidence in general for significant ongoing
particle acceleration, they represent the best cases for modelling and
hence inverse-Compton observations give the cleanest constraints on
magnetic field strengths. Early observations with {\it ROSAT} and {\it
  ASCA} allowed some field measurements to be made \citep[e.g.][]{1995ApJ...449L.149F}, but again it was
{\it Chandra} and to some extent {\it XMM-Newton} that gave us
comparatively large samples \citep{2002ApJ...581..948H,2004MNRAS.353..879C,2005ApJ...626..733C,2005ApJ...622..797K}. These provide a very
clear picture: both the detections and the non-detections of
inverse-Compton emission are consistent with a picture in which the
characteristic magnetic field in the lobes is close to, but in general
slightly below, the equipartition value. The actual field strength value depends
strongly on the lobe size and power, but might be of order 0.1--few nT
(1 to a few tens of $\mu$G) in large, bright sources, with the energy
density in electrons being perhaps an order of magnitude higher. Thus it seems that the
mechanism that ensures (approximate) equipartition in the hotspot regions may also
operate on the much larger, hundred-kpc scales of the lobes.

We can go slightly further in the best-studied systems, since the fact
that the synchrotron and inverse-Compton emission can be {\it
  resolved} with {\it Chandra} and {\it XMM-Newton} means that we are
not restricted to considering one-zone models, as first realised by
\cite{2002ApJ...580L.111I}. \cite{2005MNRAS.363..649H} showed that in fact
the magnetic field strength has to vary by a factor of up to 2
throughout the lobe of the best-studied lobe IC source, Pictor A
(Fig.\ \ref{fig-pica}), in
the sense that the field strength decreases with distance from the
hotspot. More recently, Goodger \etal\ (in prep.) have used
inverse-Compton data to argue that the strongly filamentary structure
seen in synchrotron observations of many powerful radio sources must
be a result of strong small-scale variations of the magnetic field
strength (if it were the electron energy density that was varying,
there would be a stronger correlation between the synchrotron and
inverse-Compton emission than is observed). Although this detailed
work is at the limits of what is presently achievable with X-ray
observations, it may be the only way in which the detailed
relationship between field and electrons in the lobes can be studied.

The limitation of the lobe inverse-Compton technique is that it can
only really be used to study sources in which the diffuse X-ray
emission is not dominated by thermal brems\-strahlung from the IGM; i.e.
objects in relatively poor environments. Among other things, this
means that with the exception of Cen A (see below, Section \ref{mjh:strength-progress}) there is no
inverse-Compton estimate of the magnetic field strength in the lobes
of any low-power (FRI) source, although inverse-Compton limits have
been used to constrain the particle content in such objects
\citep[e.g.][]{2008MNRAS.384.1344J,2010MNRAS.404.2018H}.

\subsection{Modelling of large-scale fields}
\label{mjh:modelling}

Analytical modelling of the dynamics of the lobes of radio galaxies
has since its beginning focussed on a pure hydrodynamical approach
\citep[e.g.][]{1974MNRAS.166..513S,1997MNRAS.286..215K} and numerical
modelling has often followed suit
\citep[e.g.][]{2003MNRAS.339..353B,2005A&A...431...45K}. To some
extent, this approach may be (retrospectively) justifiable by the
inverse-Compton observations discussed above (Section
\ref{mjh:lobe}), which tell us that magnetic fields cannot dominate
the large-scale dynamics. Nevertheless, numerical modelling involving
magnetic fields is important for three reasons; firstly, even
synthetic synchrotron mapping requires some idea of the magnetic field
strength distribution, while reproducing observations of polarization
and RM/depolarization structures (see above) is impossible without a
model of the magnetic field both in the radio source and the
environment; secondly, while the magnetic field may not dominate the
dynamics of the large-scale structures, it may still have a dynamical
role to play, for example in suppressing Kelvin-Helmholtz
instabilities; and thirdly, such modelling
in principle allows us to test assumptions about the transport of the
magnetic field structures imposed at jet generation and observed in
the parsec-scale jet (Sections 2 and 3).

The earliest simulations of lobe dynamics involving magnetic field
structure \citep{1989ApJ...342..700C,1990MNRAS.242..623M} involved
passive transport of the field, but more recently it has been possible
to numerically solve the full MHD equations for semi-realistic
density, jet speed and magnetization regimes. Some compromises have
generally been necessary to allow such simulations. For example, the
work of \cite{1999ApJ...512..105J}, \cite{2001ApJ...557..475T} and
\cite{2004ApJ...601..778T} considered electron transport and radiative
losses, but operated in a uniform environment. MHD simulations
involving realistic large-scale atmospheres, matched to the information
derived from X-ray observations
\citep[e.g.][]{2005ApJ...633..717O,2009MNRAS.400.1785G,2010ApJ...710..180O}
tend to neglect electron radiative losses and also, in earlier work,
tend to consider rather low jet density contrasts. Increased
computational power has started to allow the merger of these two
approaches \citep[e.g][]{2012ApJ...750..166M}. Key results from this
modelling to date include the broad reproduction of the lobe dynamics
expected from hydrodynamical analytic or numerical modelling; the
suppression of large-scale Kelvin-Helmholtz instabilities and of the
entrainment of external gas seen in pure hydrodynamic modelling
\citep{2009MNRAS.400.1785G}, and the
emergence of complex small-scale structure in the magnetic field and
hence the synchrotron emissivity \citep[e.g.][]{2001ApJ...557..475T}.
3D MHD simulations which can simultaneously capture realistic
large-scale dynamics and have resolution good enough to be matched to the
detailed observations of synchrotron and inverse-Compton emission in
radio galaxy lobes, so as to help with questions such as the origin of
filamentary structures seen in synchrotron emission (Section
\ref{mjh:lobe}) can perhaps be expected in the next few years.
Numerical modelling is also perhaps starting to shed some light on the
{\it origins} of the field structures on large scales; for example,
\cite{2009MNRAS.400.1785G} put in as an input condition a helical
field as observed in parsec-scale AGN jets (Section 3), and show that
this initial field is amplified by shear processes to reach much
larger values than expected from flux conservation. On the other hand,
\cite{2011MNRAS.417..382H} are able to reproduce many features of real
radio galaxy lobes with a field which is initially {\it random} (in
both the jet and the environment into which it propagates). It will be
of great interest to see whether high-resolution simulations of
large-scale jet-environment interactions can give predictions which
would allow us to distinguish between different models of the
small-scale jet magnetic field structure.

\subsection{Large-scale magnetic fields in stellar-scale outflows}
\label{mjh:stellar}

The study of the terminations and the equivalents of lobes in
stellar-scale outflows such as protostellar jets and microquasars is
much less well developed. These structures require persistent jets,
and so impulsive ejection events such as those seen in some
microquasars are not expected to produce them. Where jets are
persistent or at least recurrent on short timescales,, there is
evidence in a few well-studied microquasars for lobe-like synchrotron
emission
\citep[e.g][]{1992Natur.358..215M,2004ASPRv..12....1F,2005A&A...434...35H}
and even both lobes and hotspots \citep{2010MNRAS.409..541S}, as well
as indirect evidence for the existence of lobes whose synchrotron
emission is not seen in some other cases \citep{2005Natur.436..819G}.
The lack of lobes in general in these systems has been attributed to
the low-density environments in which they reside
\citep{2002A&A...388L..40H}. Even when found, though, these lobes are
generally faint, so that even polarization measurements are generally
not possible, and techniques such as magnetic field strength
measurements by inverse-Compton emission are completely out of the
question, even setting aside the complex technical challenges of doing
so in the presence of an often strong central X-ray source and of a
complex and poorly constrained photon field. Authors wishing to
estimate magnetic field strengths use the equipartition assumption,
despite its limitations (Section \ref{mjh:strength}).

Turning to protostellar outflows, models for these systems fall into
two general categories: a jet-driven bow shock picture analoguous to
the dynamics of FRII radio galaxies, and a wind-driven shell picture
in which the molecular gas is driven by an underlying wide-angle wind
component, such as is given by the X-wind \citep{Cabrit/etal:1997}. A
survey of molecular outflows by \cite{Lee/etal:2000} found that both
mechanisms are needed in order to explain the full set of systems
observed. The advent of Spitzer observations of protostellar jets has
led to spectacular advances in our understanding of jet morphology,
shock structure, and jet termination. For example, the jet in the
classic T-Tauri star system HH46/47, imaged in the infrared with high
resolution techniques (down to sub-arcsec levels), is bright and
highly collimated and has both a wide-angle outflow cavity as well as
a collimated jet that extends through all scales from the protostellar
scale up to the termination point on the bow shock
\citep{Velusamy/etal:2007}, showing a morphology which is remarkably
similar to those of powerful radio galaxies. The jet is well
collimated with a width of 1-1.5$''$ and length of 10$''$ (4500 AU).
One of the most prominent features are bright ``hotspots' that are
coincident with the head of the jet and appear to be analogous to the
hotspots of radio galaxies. Cool molecular gas is detected in the
wide-angle flow that surrounds this jet. Similarly, observations of
the jet Cep E associated with a more massive forming star (with up to
4 $M_{\odot}$), using the same techniques, show shocked molecular
hydrogen $H_2$ emission --- which requires the existence of a magnetic
precusor (C-shocks) in order to be excited \citep{Velusamy/etal:2011}.
Here again, the hottest emission comes from atomic/ionic gas produced
at the hot spot at the bow shock of the jet. However, despite this
important evidence for magnetization of the shocked region, and unlike
the AGN case, there is as yet little direct evidence for the role of
magnetic fields in the production of ``hotspots'' and lobes, since
there are few observational diagnostics of field strength in this
regime and synchrotron radiation is seldom observed. In the case of
protostellar jets, the state of the art, as discussed above
\citep{Carrasco-Gonzalez/etal:2010} is the detection of radio
polarization, confirming a synchrotron origin for some of the detected
radio continuum, but this is a detection of the jet, rather than of a
lobe-like structure, and here again equipartition assumptions must be
used to estimate a magnetic field strength. Next-generation radio and
X-ray facilities will be required for the study of these systems to
catch up with that of extragalactic large-scale lobes.

\section{New directions}

\subsection{Magnetic field strengths and directions in protostellar jets}

The Expanded VLA (EVLA) and the extended Multi-Element Radio Linked
Interferometer (e-MERLIN) will have an order of magnitude more
sensitivity than their predecessors. Much larger parts of the radio
spectrum will be measured as a consequence, and this will have a huge
impact on our ability to measure synchrotron radiation from
protostellar systems. The one detected source HH80-81 is one of the
most powerful TTS jets known. If the weaker jets turn out to have
measurable synchrotron emission, this will herald a new era in the
study of jets in general, and of their magnetic field structure in
particular.

A second major advance in understanding the launch of jets will come
with the ALMA observatory. ALMA observations will allow one to resolve
the 1-AU scale of nearby protostellar discs for the first time. The
important point is that ALMA will also be able to measure linearly
polarized emission. It is expected that this will allow one to map the
structure of magnetic fields on the disc, and, in this way, to study the
structure of the disc field and its connection to the launch
mechanism.

The bottom line is that the era of real magnetic field measurements in
protostellar systems is very near --- and this will help revolutionize
our knowledge of how jets are launched and collimated.

\subsection{Parsec-scale magnetic fields in AGN: Physics of the
launching, propagation and confinement of the jets}

Observations with the Very Long Baseline Array over the past decade have
made the importance of high-resolution, multi-frequency, polarization-sensitive
radio studies with VLBI abundantly clear. The availability of spectral
information can help identify locations of particle re-acceleration and
low-frequency absorption in the jets, and the Faraday rotation distribution
can provide information about both the thermal gas in the vicinity of the
AGN and the intrinsic magnetic fields of the jets themselves. The question
of whether there is an appreciable helical or toroidal component is of
crucial importance for our understanding of the launching and confinement
of the jets, and will be a key aspect of future observational studies.
Multi-frequency polarization VLBI observations of more AGNs are very much
needed for these reasons.

Faraday-rotation studies on parsec scales with high resolution are of
particular interest; however, work of this kind has been hindered  by
the fact that increasing resolution requires observing at higher frequencies,
whereas increasing the sensitivity to Faraday rotation requires observing at
lower frequencies. One promising way to overcome this limitation is through
space VLBI observations at centimetre wavelengths. The currently flying
RadioAstron mission is likely to be of only limited use for this purpose,
since it has limited imaging capability due to the elongated orbit of the
space antenna.  The Japanese-led VSOP2 mission would have provided much
better imaging capability, but, unfortunately, this mission has recently been
cancelled. It will be important to ensure that any other future space-VLBI
missions include frequencies low enough (8, 5 and possibly even 1.6 GHz)
to make these power tools for high-resolution Faraday-rotation studies.
On the other hand, polarization observations at 3~mm-1.3~cm sampling
subparsec scales could also prove valuable for future Faraday-rotation
studies, since these will probe regions with high electron densities and
high magnetic fields in the innermost jets, where very high Faraday rotations
may be reached.

Another area in which future VLBI observations can play an important role is in
relatively low-frequency multi-frequency observations with global arrays
such as the VLBA. Such observations can help link up the parsec scales
studied this far with VLBI and the kpc scales that will be
studied with high sensitivity using the EVLA and e-MERLIN. One such project
is a recently completed set of four-wavelength 18-22~cm polarization VLBA
observations of the 135 AGNs in the main MOJAVE sample, which typically provide
information about the jet structures out to scales of tens of parsec.
Observations of additional AGNs at these and longer wavelengths would
be of considerable interest. It is thus extremely important to develop
techniques for accurate intensity, polarization and polarization position
angle calibration of such long-wavelength VLBI data.

\subsection{Kiloparsec-scale AGN: new approaches to field direction
  and strength}

\label{mjh:direction-progress}

The EVLA, and other next-generation radio telescopes with observing
bandwidths of order GHz rather than tens of MHz, will have a great
impact on the study of Faraday rotation in the large-scale polarized
components of radio galaxies. Rather than requiring many monochromatic
multi-frequency observations, it will in many cases be possible to
make a good measurement of $RM$ and depolarization with a single
broad-band observation. This should allow the results obtained from
detailed studies of bright objects in rich environments \citep{1987ApJ...316..611D,1996cyga.book..168P} to be generalized to more typical
systems. The sensitivity and resolution provided by the EVLA and
e-MERLIN will allow the detailed, quantitative study of polarization
structures in jets (Section \ref{mjh:kpc-jet}) to be extended to more
objects, to smaller scales (connecting 100-pc and pc-scale jets) and
possibly to the large-scale jets in FRII sources. At lower
frequencies, instruments such as LOFAR with the capability to measure
polarization sensitively at low frequencies may allow us to probe the
$RM$ due to very tenuous, large-scale gas seen in front of the
polarized outer regions of giant radio galaxies, giving constraints on
$n_e$ and $B$ on these scales that are currently unobtainable in any
other way.
\label{mjh:strength-progress}

Progress in inverse-Compton measurements of $B$ will also have to come from
new observational developments. Currently one problem with the method
is the extrapolation between the properties of the $\gamma \approx
1000$ electrons responsible for inverse-Compton scattering CMB photons
into the X-ray band and the $\gamma \approx 10^4$ electrons
responsible for radio synchrotron emission at GHz frequencies. This
situation will be improved significantly with the advent of
next-generation low-frequency radio telescopes such as LOFAR. We can
also hope, with less certainty, for imaging missions working at harder
X-ray energies, such as NuSTAR, which will not only probe somewhat higher-energy
electrons but, by operating above the cutoff energies for thermal
bremsstrahlung, will make it much easier to study inverse-Compton
emission from sources in rich environments. Another ambiguity in
models is the correct modelling of the very low-energy electron
population: as shown by \cite{2008MNRAS.385.2041C}, ALMA may in principle be
able to detect the Sunyaev-Zel'dovich effect from radio galaxy lobes,
which is a very strong probe of the total number of relativistic
electrons and hence of the low-energy cutoff (although see \citealt{2008MNRAS.388..176H} for some caveats). Finally, inverse-Compton studies at
very high energies have some promise. {\it Fermi} recently made the
first detection of what appears to be inverse-Compton scattering of
the CMB to $\sim$ GeV energies, in the nearby, and therefore very
large in angular size, radio galaxy Cen A \citep{2010Sci...328..725A}: taken
at face value the detection implies a magnetic field strength in the
lobes that is again close to the equipartition value. {\it Fermi} can
resolve only a very few nearby and (necessarily) atypical radio
galaxies, but even limits from this process may be useful. At higher
energies still, existing detections with TeV imaging instruments such
as HESS and MAGIC place limits on inverse-Compton scattering
(predominantly of starlight: \citealt{2003ApJ...597..186S}) by the TeV electrons
in the jets of low-power radio galaxies: for example, existing HESS
detections of Cen A, whose synchrotron properties are very well known
(e.g.\ \citealt{2007ApJ...670L..81H} and refs therein) already require that
the magnetic field cannot be too much lower than the equipartition
value, although detailed modelling is complex \citep{2011MNRAS.415..133H}. The resolution and sensitivity that will be achieved with
instruments such as the planned Cerenkov Telescope Array (CTA) will
actually allow imaging of the TeV inverse-Compton jets in objects such
as Cen A, if $B \approx B_{\rm eq}$, and so will permit mapping of the
$B$-field in an X-ray synchrotron jet for the first time. In the very
long term, X-ray telescopes with very high sensitivity (e.g., the
currently planned {\it
  Athena} mission) may allow us to measure inverse-Compton emission
from the lobes of Galactic microquasars and synchrotron-emitting
protostellar jets (Section \ref{mjh:stellar}).

\section{Conclusion}

In conclusion, this review has covered a vast amount of observational,
theoretical and computational phase space with the purpose of drawing
out and analyzing the importance of magnetic fields for the origin and
structure of astrophysical jets. The underlying engines for all of
these systems involves rotating magnetized objects such as stars,
black holes, and accretion discs. Relativistic and even general
relativistic MHD leads to similar conclusions about jet dynamics to
those deduced (and in some cases observed) in protostellar systems.
While measurements of densities, temperatures, and velocities is
straightforward in the latter type of system, magnetic measurements
are very difficult. Exactly the opposite is the case for AGN jets.
Encompassing these two complementary worlds lies the complete picture
of astrophysical jets which is still in a very exciting stage of
exploration. It will be important therefore to keep probing and
extending our ability to measure magnetic field strengths and
geometries in these diverse systems, since these provide insight into
the origin of one nature's most marvellous creations.

\section*{Acknowledgements}

We thank Andr\'e Balogh and the organizers of an ISSI workshop on
``Large Scale Magnetic Fields in the Universe'' for their invitation
to review this topic. This paper has benefited from helpful comments
from an anonymous referee. The research of REP is supported by a
Discovery Grant from the National Science and Engineering Research
Council of Canada. MJH acknowledges support from the Royal Society and
the UK Science and Technology Funding Council.

\bibliographystyle{aps-nameyear}      
\bibliography{mjh,dcg,ralph}   

\end{document}